\documentclass[12pt]{iopart}

\usepackage{iopams}
\expandafter\let\csname equation*\endcsname\relax
\expandafter\let\csname endequation*\endcsname\relax
\usepackage[usenames,dvipsnames]{xcolor}
\colorlet{color1}{NavyBlue}
\usepackage[colorlinks=true,allcolors=color1]{hyperref}
\usepackage{amssymb}
\usepackage{physics}
\usepackage{dsfont}
\usepackage{graphicx}
\usepackage{xfrac,overpic}
\usepackage[utf8]{inputenc}
\usepackage{enumitem}
\usepackage{soul}

\begin{document}

\title{Black hole inner horizon evaporation in semiclassical gravity}

\author{Carlos Barceló$^1$, Valentin Boyanov$^{2}$, Raúl Carballo-Rubio$^{3}$ and Luis J. Garay$^{2,4}$}

\address{$^1$Instituto de Astrofísica de Andalucía (IAA-CSIC), Glorieta de la Astronomía, 18008 Granada, Spain\\
	$^2$Departamento de Física Teórica and IPARCOS, Universidad Complutense de Madrid, 28040 Madrid, Spain\\
	$^3$Florida Space Institute, 12354 Research Parkway, Partnership 1, Orlando, FL 32826-0650, USA\\
	$^4$Instituto de Estructura de la Materia (IEM-CSIC), Serrano 121, 28006 Madrid, Spain}
\ead{carlos@iaa.es, vboyanov@ucm.es, raul.carballorubio@ucf.edu, luisj.garay@ucm.es}
\vspace{10pt}
\begin{indented}
	\item[]December 2020
\end{indented}

\begin{abstract}
	In this work we analyse the backreaction of a quantum field on a spherically symmetric black hole geometry with an inner horizon, i.e. an internal boundary of the trapped region. We start with a black hole background with an inner horizon which remains static after its formation. We quantise a massless scalar field on it and calculate its renormalised stress-energy tensor in the Polyakov approximation. We use this tensor as a source of perturbation on top of the background spacetime. We find that the inner horizon has a tendency to evaporate outward much more quickly than the outer one evaporates inward through the Hawking effect. This suggests a revised picture of a semiclassically self-consistent evaporation in which the dominant dynamical effect comes from the inner horizon, the cause of which can be seen as an interplay between the well-known unstable nature of this horizon and a locally negative energy contribution from the quantum vacuum. We also look at backreaction on backgrounds which resemble gravitational collapse, where the inner horizon moves toward the origin. There we find that, depending on the nature of the background dynamics, horizon-related semiclassical effects can become dominant and invert the collapse.
\end{abstract}

\maketitle

\section{Introduction}

Quantum field theory in curved spacetimes tells us that an initially empty quantum vacuum state can evolve, through the dynamics of the background spacetime, to have a content of particles, with a corresponding energy and momentum \cite{BD,Wald1995}. This result has lead to crucial developments in the field of gravitational physics and cosmology, the most notable of which is perhaps Hawking's analysis of the evaporation of black holes \cite{Hawking1975}. It shows that the backreaction of the quantum vacuum on spacetime can lead to classically forbidden results, such as extracting mass from a black hole. Further, it has since been shown that the evaporation of black holes has certain universal and local (in both time and space) characteristics, the only condition for which is that the black hole be formed by gravitational collapse \cite{Wald1995,Barceloetal2010}.\par
It is therefore generally accepted that black holes should evaporate. More precisely, this means that their trapped region should gradually reduce in size from the outside in, eventually revealing what is at their core, where the simplest, spherically symmetric model tells us that there is either a singularity, or some nucleus which can only be described with a complete theory of quantum gravity. However, this picture is missing an essential ingredient, one which is present whenever the black hole structure is considered with sufficient generality: the inner horizon, i.e. the inner bound of the trapped region, below which some causal observers are free to stop their descent.\par
When a trapped region forms during gravitational collapse, it usually has an outer and inner boundary, comprised of two apparent horizons. The outer one classically only moves outwards, and does so as more matter is accreted, while the inner one moves inwards, either tending to a final position at a finite radius (if e.g. the black hole has any angular momentum or electric charge, or forms a regular core by violating the strong energy condition) or collapsing all the way to the origin and creating a spacelike singularity. An inner horizon appears in all realistic black hole configurations, and one may therefore wonder whether there is some semiclassical effect associated with its presence, just as the Hawking effect is related to the presence of the outer horizon. Models of evaporation of trapped regions with an inner horizon have indeed been considered, but they take evaporation as an ad hoc ingredient, implicitly assuming that it still mainly occurs from the outside, and that the inner horizon just waits around for the outer one to eventually come to it (see e.g. \cite{Hayward,Frolov2017}). The present work is, to our knowledge, the first attempt at analysing the behaviour of the inner horizon under semiclassical perturbations without this bias.\par
The evolution of the geometry in the vicinity of the inner horizon is not a straightforward matter even in classical physics, as mild perturbations can lead to the well-known mass inflation instability \cite{PoissonIsrael89}. The backreaction from this instability considerably alters the geometry close to the inner horizon \cite{Brady1995,Marolf2012}, though, in keeping with causality and energy positivity, it only allows this horizon to move inwards.\par
In this work we perform an analysis of the backreaction on this region brought about solely by semiclassical sources, considering that all classical ones are part of the background. Our goal is to check the validity of the standard assumption that evaporation of the trapped region primarily occurs from the outside, and that the inner horizon is left only to its classical evolution. In other words, we want to analyse the dynamics of the inner horizon when sourced by the energy present in the quantum vacuum after a black hole is created through gravitational collapse. To this end, we first construct a toy model geometry in which a spherical black hole with an inner horizon forms by the collapse of a null thin shell. As a source of backreaction we use the renormalised stress-energy tensor (RSET) of a quantum scalar field in the Polyakov approximation \cite{Fabbri2005}, which is known to capture the essential characteristics of evaporation when used at the outer horizon \cite{DFU,Fabbri2005}. Whether this approximation gives the complete picture of semiclassical effects at the inner horizon is not clear, as no complete calculation of the RSET in 3+1 dimensions for this region is available, the closest being e.g. the asymptotic analysis in \cite{Wald2019}. Nevertheless, we believe a detailed calculation using the Polyakov approximation is worthwhile, at the very least in order to get a first glimpse of what backreaction at early times may look like in a tractable toy model.\par
Our model is by construction simple enough for most of the calculations to be analytical, and thus serves as an example in which the origin of backreaction and the evolution of the subsequent self-consistent solution are clear enough. In more general scenarios we find that the problem completely lacks the universality of late-time behaviour present in its outer horizon counterpart, and it is therefore hard to establish any general conclusions for the complete evolution of such spacetimes. However, we see in our model that the semiclassically induced dynamics of the inner horizon are important in the picture of evaporation as a whole, which is highly indicative that this is the case in general.\par
The result for our model is that backreaction causes the inner horizon to move, much as it does the outer one, and more quickly at that. Particularly, the initial tendency in the evolution is for the inner horizon to move outward much more quickly than the outer one moves inward under standard Hawking evaporation. Although we do not yet have a complete picture in which both horizons meet and the trapped region disappears (as our calculation is only valid for the initial stages of evaporation), this result strongly suggests a revised picture in which the semiclassical dynamics of the inner horizon are at least as important as those of the outer one, if not more.\par
The local nature of our analysis and the simplicity of the calculations allow us to extend our model to include dynamical inner horizons, which on the one hand lets us obtain a better approximation to the complete semiclassically self-consistent dynamics (although once again only in a limited time interval). On the other hand it allows us to include backgrounds in which such dynamics are classically present, as is the case in most scenarios of gravitational collapse (e.g. the Oppenheimer-Snyder model \cite{Oppenheimer1939}, see discussion below). Reproducing the dynamics of such scenarios, we find that there are cases in which the semiclassical contribution to the equations can become dominant, going as far as inverting the collapse, through only the horizon-related terms of the RSET.\par
The formalism we build can be used to find the semiclassical effects in many dynamical black hole causal structures, and we have far from exhausted its applicability. For instance, one important issue that is still not completely solved is the superposition between this semiclassical evaporative effect and classical mass inflation, which seems to work in the opposite direction \cite{Brady1995}. We note, however, that there are strong indications that at late times it is the semiclassical effect which dominates \cite{Wald2019}, as we will discuss in more detail below.\par
The structure of our work is the following. In section \ref{innerh} we give an overview of black-hole geometries with inner horizons, both static and dynamical ones, for which the conclusions of our analysis may be relevant. For the static (or nearly static) case we also review the instabilities inherent to such spacetimes due to the presence of a Cauchy horizon. In section \ref{static} we present our toy model for the formation of a black hole with an inner horizon, we calculate the RSET for the ``in" vacuum state, and we solve the first order perturbation to the Einstein equations, leading to our result of an evaporating trapped region. In section \ref{dynamic} we extend our model to one with a moving horizon. Here we obtain an approximation for the semiclassically self-consistent dynamics and also analyse geometries with classically dynamical horizons. In the final section we summarise the main conclusions of this work and briefly discuss their possible implications for the black holes observed in our universe.

\section{Inner horizon in static and dynamical black holes}\label{innerh}

Throughout this work, we use the term \textit{inner horizon} to refer to the innermost (marginally) trapped surface of a black-hole trapped region, following the definitions in \cite{Hayward1993}. The most well-known geometries with an inner horizon are the Kerr and Reissner-Nordström solutions \cite{Wald1984}. It is also a common feature of regular (singularity-free) black holes (see e.g. \cite{Ansoldi} for a review and \cite{Carballo-Rubio2019,Carballo-Rubio2019b} for more recent results). Additionally, a dynamical (short-lived) version of this horizon appears in almost any black hole formation scenario. The toy model geometries we use throughout this work are built to reflect the causal structure of these spacetimes, so to begin with we will quickly review them. We also dedicate part of this section to a bibliographic summary of results on the classical and semiclassical mass inflation instability, the latter being closely related to backreaction from the RSET on the inner horizon.\par

\subsection{Classical geometries with an inner horizon}

Our main interest lies with geometries which are asymptotically flat in the past. Aside from being a good initial description for spacetimes with dynamically formed black holes, this also allows us to avoid ambiguities in the definition of an initial quantum vacuum state. Therefore, we will look at spacetimes in which a black hole forms from an initially dispersed distribution of matter. For simplicity, we restrict our study to spherically symmetric configurations. The geometries we will consider will generically have the form
\begin{equation}\label{metric1}
ds^2=-f(r)dt^2+\frac{1}{f(r)}dr^2+r^2d\Omega^2,
\end{equation}
where $f(r)$ will be referred to as the \textit{redshift function}, and $d\Omega^2=d\theta^2+{\rm sin}^2\theta\mathop{d\varphi^2}$ is the line element of the unit two-sphere. In principle, one could consider more generic geometries with two degrees of freedom and, in fact, we will do this when considering perturbations of these backgrounds. However, we will see explicitly that the geometries above are general enough to describe the backgrounds we are interested in.

\subsubsection{Charged and regular black holes}

For example, a charged black hole metric is given by
\begin{equation}\label{RN}
f(r)=1-\frac{2M}{r}+\frac{Q^2}{r^2},
\end{equation}
with $M$ being the mass and $Q$ the electric charge. The outer and inner horizons are static and have radial positions $r_\pm=M\pm\sqrt{M^2-Q^2}$, as long as $M>Q$ (if $M=Q$ we get the extremal configuration, while if $M<Q$ we get a naked singularity, though we will not discuss these cases here).\par
\begin{figure}
	\centering
	\includegraphics[scale=.7]{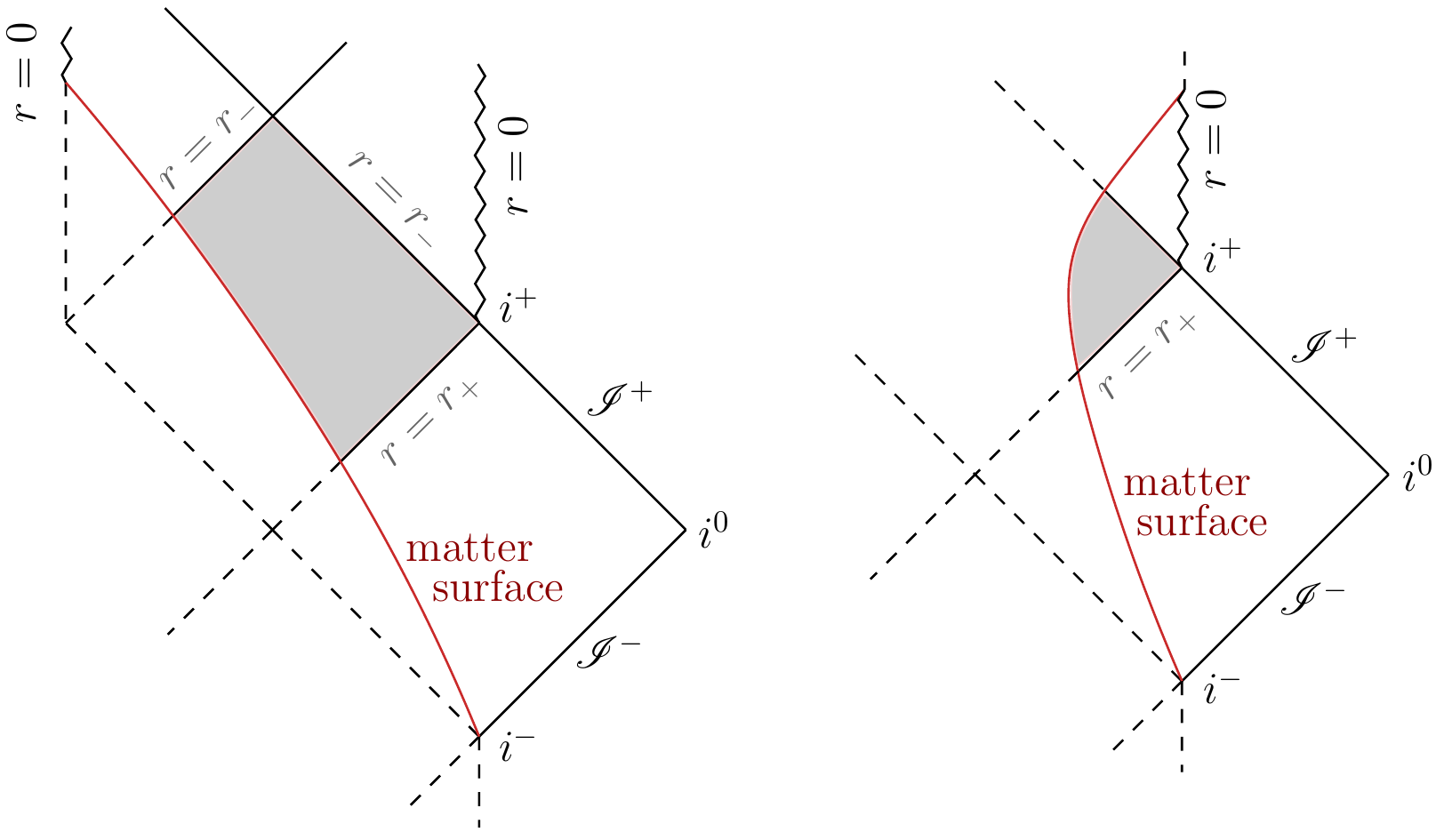}
	\caption{Causal diagrams for the formation of a charged black hole, and their analytical extensions. On the left diagram matter collapses to form a singularity before crossing the Cauchy horizon. On the right, matter crosses the Cauchy horizon before forming a singularity. The shaded regions are the exterior parts of the trapped regions in each case. These diagrams also represent the formation of a regular black hole if we just substitute the singularities with regular points.}
	\label{f11}
\end{figure}

For a geometry in which a charged black hole forms from a collapse of a spherical matter distribution, this exterior geometry is matched with the matter-filled interior along the trajectory of the spherical surface, as shown in figure \ref{f11}. There are two topologically distinct ways for this matching to occur. The left diagram of the figure shows the first possibility: matter generates both an outer and inner horizons, which maintain constant radial positions $r_\pm$ once its surface gets past them, from where it continues its descent and a timelike singularity is formed at a finite time within the initial universe. The right diagram shows the second possibility: matter generates an outer horizon but then crosses the Cauchy horizon it forms at the same time its surface crosses the inner radius $r_-$, before collapsing all the way to the origin. This latter case, although more peculiar from a physical point of view, is in fact the solution obtained for the collapse of a charged pressureless thin shell \cite{Boulware1973,Wald1984}.\footnote{Technically there is a third possibility in which the surface first crosses the inner radius $r_-$ and then turns around to cross the Cauchy horizon instead of collapsing to the origin. The conclusions we obtain from the second scenario apply directly to this third one, so we will not consider them separately.}\par

\begin{figure}
	\centering
	\includegraphics[scale=.6]{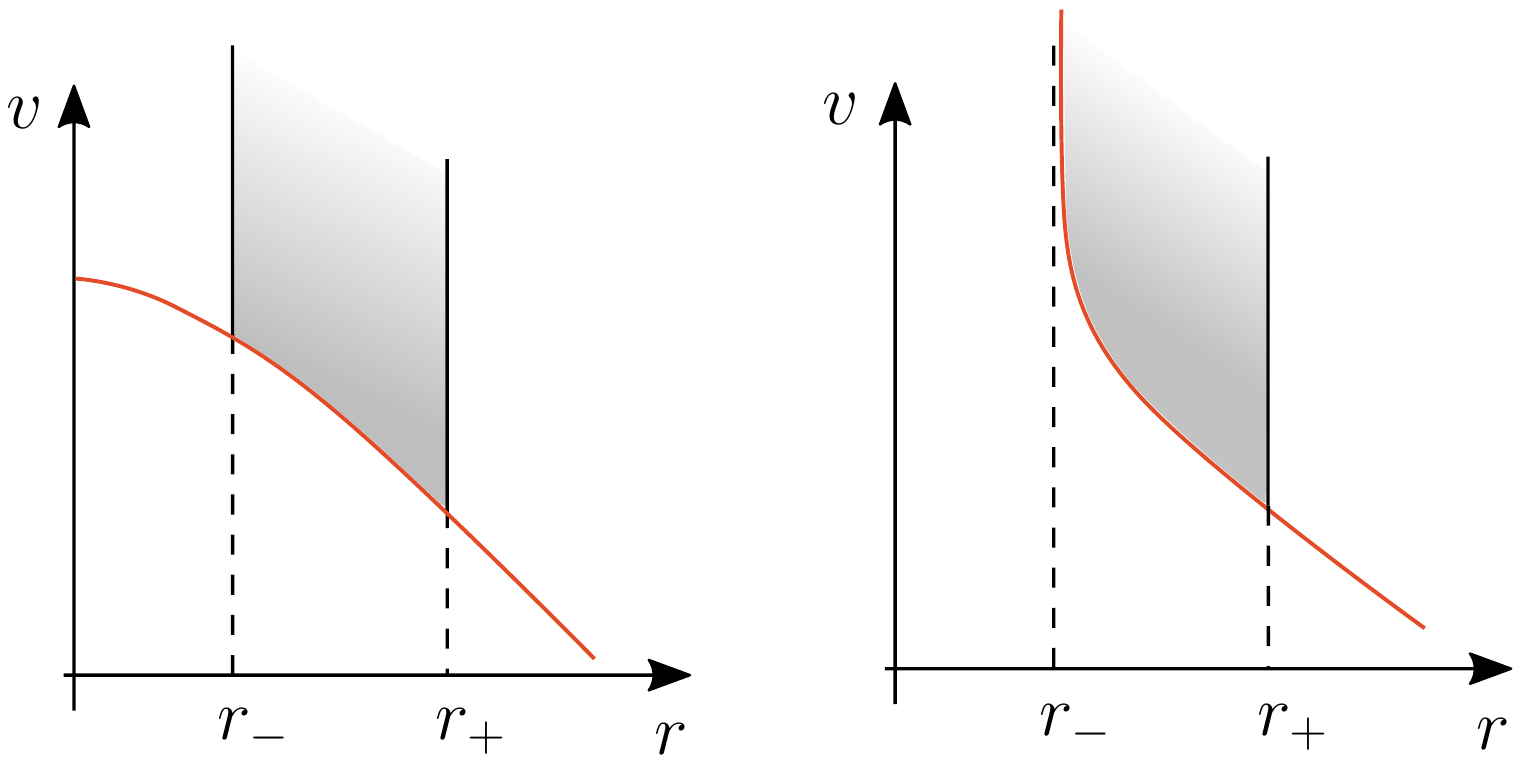}
	\caption{Two possible trajectories of a collapsing charged shell of matter shown in advanced Eddington-Finkelstein coordinates. The plot on the left shows the possibility of matter collapsing to the origin at finite $v$, i.e. within the confines of the initial universe, crossing its inner horizon but not its Cauchy horizon. The plot on the right shows the possibility of the shell remaining outside the inner radius $r_-$ but tending toward it asymptotically in $v$, eventually crossing the Cauchy horizon it generates (simultaneously leaving the portion of spacetime covered by the $v$ coordinate).}
	\label{f12}
\end{figure}

It is interesting to note that these two scenarios are actually quite similar while within the confines of the initial universe, that is, as far as the dynamics of the inner horizon are concerned. To see this, let us consider the simplest case in which the collapsing distribution of matter is a thin shell. Once the shell crosses its outer horizon at $r_+$ a trapped region is formed, the upper bound of which is the horizon and the lower bound is dynamical and coincides with the matter shell itself, up until the latter crosses $r_-$. The two scenarios are plotted in advanced Eddington-Finkelstein coordinates $(v,r)$ in figure \ref{f12}. It is easy to see that the trajectory of the inner horizon in the second case is only slightly different from the one in the first, a difference which quickly tends to zero as $v$ increases. The main difference between the two cases is then whether or not there is a singularity at the origin formed in finite advanced time.\par

These characteristics are the same when analysing the formation of a regular black hole rather than a charged one, the only difference being the absence of a singularity at $r=0$ when the collapse is finished. In this work, we will in fact work with geometries in which a regular black hole is formed, as defining a quantum vacuum around a timelike singularity is a rather ambiguous process. Nevertheless, our results regarding the semiclassical dynamics of the inner horizon will be equally applicable to the charged black hole, so long as this horizon remains outside the range of causal effect of the singularity.

\subsubsection{Oppenheimer-Snyder collapse}

\begin{figure}
	\centering
	\includegraphics[scale=.8]{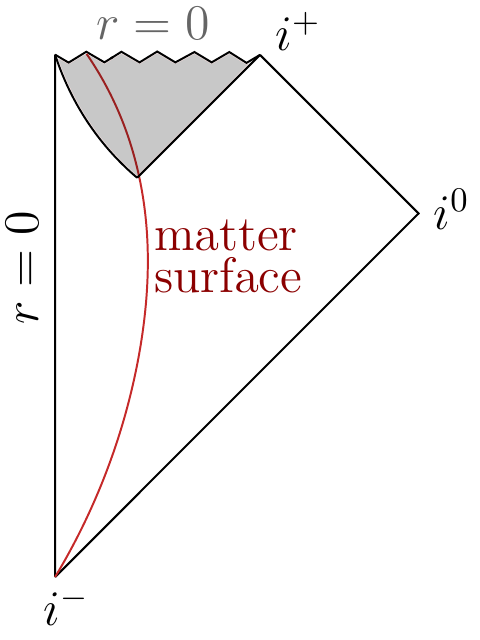}
	\caption{Causal structure of the Oppenheimer-Snyder-type geometry. The red curve represents the surface of the collapsing sphere of matter. The shaded region is trapped, and the inner horizon is its left border.}
	\label{fig23}
\end{figure}

When a Schwarzschild black hole forms from gravitational collapse, the trapped region is generally formed before the singularity and has both an outer and inner apparent horizons. The outer horizon is either stationary or moves outward until all the collapsing matter has crossed it. The inner horizon moves inward and reaches the origin as the singularity is formed. The causal diagram in figure \ref{fig23} illustrates this (see \cite{Bengtsson2013} for a more detailed discussion).\par

Here we will briefly remind the reader of one example of such a collapse: the Oppenheimer-Snyder model \cite{Oppenheimer1939}. This model represents the gravitational collapse of a pressureless, homogeneous ball of dust. Its geometry can be constructed by matching a section of a closed Friedman-Robertson-Walker universe for the interior with a patch of Schwarzschild spacetime for the exterior. The metric for the interior section is then most conveniently expressed in cosmological coordinates,
\begin{equation}
ds^2=a^2(\tau)\left(-d\tau^2+d\chi+{\rm sin}^2\chi d\Omega^2\right),
\end{equation}
where $a(\tau)$ is the conformal factor, which has the form
\begin{equation}
a(\tau)=\frac{a_0}{2}\left(1+{\rm cos}\,\tau\right).
\end{equation}
The collapse starts at $\tau=0$ and ends at $\tau=\pi$. The coordinate $\chi$ goes between $0$ and $\chi_0<\pi/2$. The two constants $a_0$ and $\chi_0$ are related to the initial conditions through
\begin{equation}
a_0=\sqrt{\frac{r_0^3}{2M}},\qquad {\rm sin}^2\chi_0=\frac{2M}{r_0},
\end{equation}
where $r_0$ is the initial radius of the ball and $M$ is its mass.\par
\begin{figure}
	\centering
	\includegraphics[scale=.5]{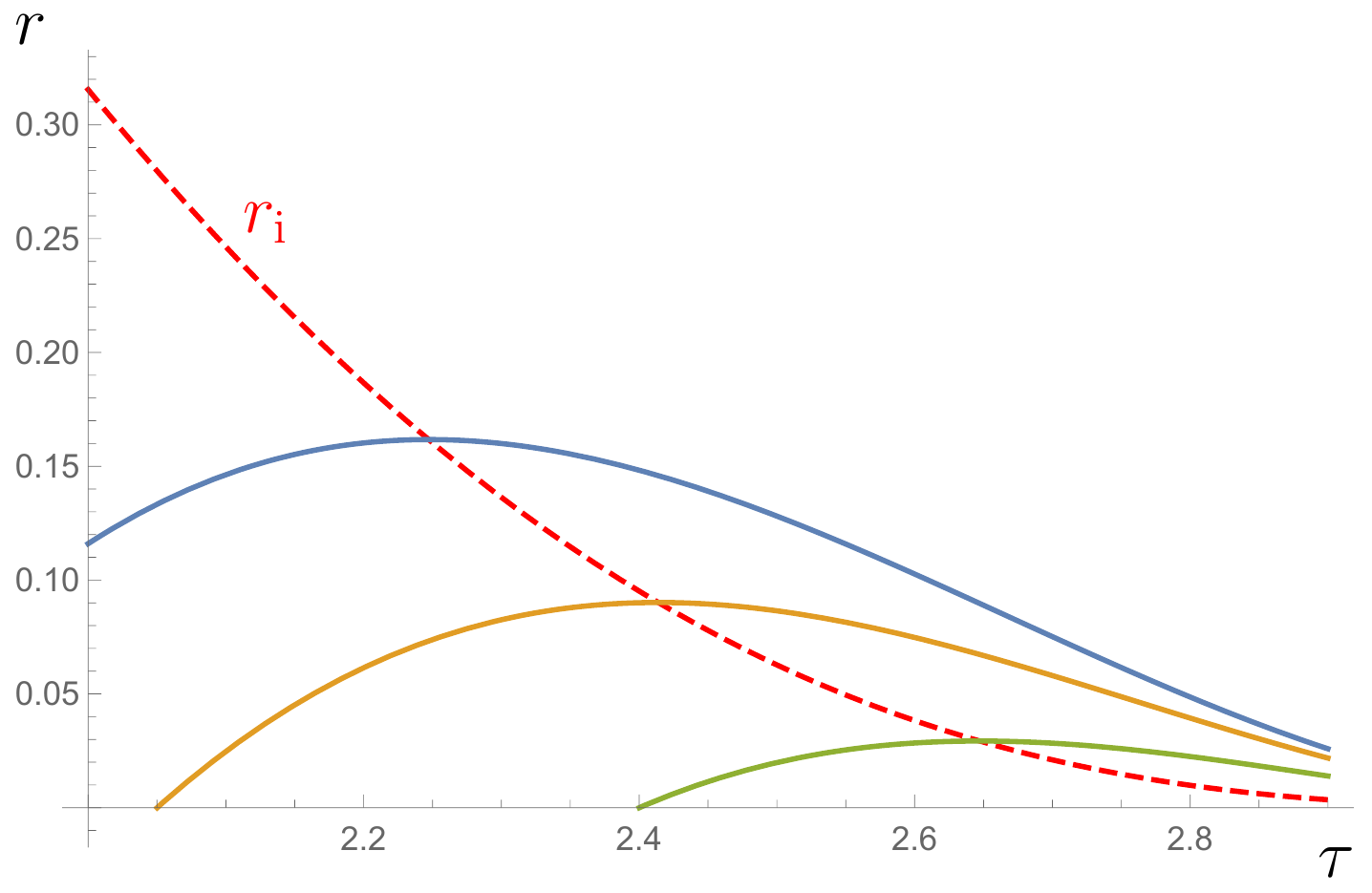}
	\caption{Outgoing null trajectories in the Oppenheimer-Snyder interior region. The dashed line represents the dynamical inner horizon, beyond which even these null rays start moving inward.}
	\label{fig22}
\end{figure}

Now, it is clear that in this interior region causal trajectories are seemingly allowed to move outward throughout the whole collapse, even when the surface has already crossed the Schwarzschild radius of the external geometry. However, when considering whether this movement is actually in the direction of increasing radius, we must take into account that it is relative to the collapsing matter distribution. The radius, as defined by the surface area of spheres in sections of constant $\tau$ and $\chi$, is given by
\begin{equation}
r(\tau,\chi)=a(\tau){\rm sin}\,\chi.
\end{equation}
If we consider the quickest causal outward movement, i.e. outgoing radial null geodesics, given by $\tau=\chi+U$ with $U$ a constant in the range $(0,\pi)$, their radial positions are given by
\begin{equation}\label{OSnull}
r=\frac{a_0}{2}(1+{\rm cos}\,\tau){\rm sin}(\tau-U),
\end{equation}
with $\tau$ ranging between $U$ and either $U+\chi_0$, if the latter is less than $\pi$ (in which case the light ray escapes from the surface into the Schwarzschild region), or up to $\pi$ if the opposite inequality is satisfied (in which case the ray remains in the interior until it falls into the singularity). What we are interested in is the inner apparent horizon, where outgoing light rays switch from going in a direction of increasing $r$ to one of decreasing $r$. In terms of the parameter $\tau$ this can be simply obtained by looking for the spot where the derivative of \eqref{OSnull} with respect to it becomes zero. The trajectory of the inner horizon is thus described by the timelike curve
\begin{equation}\label{OSri}
r_{\rm i}(\tau)=\frac{a_0}{2}(1+{\rm cos}\,\tau){\rm sin}\left(\frac{\pi-\tau}{2}\right).
\end{equation}
This relation is only valid when $r_{\rm i}<r_{\rm s}$, with $r_{\rm s}(\tau)=a(\tau){\rm sin}\,\chi_0$ being the position of the surface, which implies that it is valid for $\tau\in(\pi-2\chi_0,\pi)$. At the lower bound of this interval we have $r_{\rm i}=r_{\rm s}=2M$, i.e. the inner horizon is formed at the same moment as the outer horizon, it stays within the matter distribution and reaches the origin when the singularity is formed. Figure \ref{fig22} shows part of the trajectory of the inner horizon $r_{\rm i}(\tau)$, superposed with the trajectories of a few outgoing light rays.\par

We thus see an example of the fact that gravitational collapse in almost all its forms is accompanied by an inner horizon, however briefly. But even this brief existence may create a sufficient environment for the non-local terms of the RSET, usually magnified at horizons, to manifest. This may be so here especially due to the fact that the time scale in which such effects become important is usually related to the surface gravity of the horizon $k_1$ (a term we use to refer to the radial slope of the redshift function, generalised from its definition in the presence of Killing vector fields), and this surface gravity tends to a divergence as the singularity is approached, as we will see. It will therefore be particularly useful for our later analysis to look at how $r_{\rm i}$ approaches the origin, and how $k_1$ diverges there. Expanding \eqref{OSri} around the singularity ($\tau=\pi$) we get the leading order term
\begin{equation}
r_{\rm i}=\frac{a_0}{8}(\pi-\tau)^3+\cdots.
\end{equation}
In the next sections we will use the exterior advanced Eddington-Finkelstein coordinates to describe the whole geometry (see eq. \eqref{metric-vr} below), so we note that this expansion in these coordinates becomes
\begin{equation}\label{OSv}
r_{\rm i}\propto(v_{\rm s}-v)^{3/4}+\cdots,
\end{equation}
where $v_{\rm s}$ is the instant of singularity formation. To see how the surface gravity diverges, first we need to define this quantity more precisely. Following the procedure from our previous work \cite{Barcelo2020}, we will use the \textit{generalised redshift function} $F(v,r)$, which goes on the right-hand side of the equation for outgoing null radial geodesics,
\begin{equation}
\frac{dr}{dv}=F(v,r).
\end{equation}
We define the surface gravity at a dynamical horizon as the absolute value of the slope in the radial direction of $F(v,r)$ at the horizon.\footnote{In the static case this definition differs slightly from what is usually referred to as surface gravity \cite{Wald1984}, coinciding only for metrics of the type \eqref{metric1}; perhaps a better term for this quantity would be \textit{thermal surface gravity}, but since we use metrics of the type \eqref{metric1} in the rest of this work it would be an unnecessary complication.} With this definition we obtain the divergent expression for the surface gravity
\begin{equation}\label{OSk}
k_1\propto\frac{1}{r_{\rm i}\,\text{sin}\,\tau}\propto\frac{1}{v_{\rm s}-v}.
\end{equation}
It is fairly easy to understand the origin of this expression: the divergence would go as $1/r_{\rm i}$ if the profile of $F$ in the $r$ direction were a straight line starting from a fixed (positive) point at the origin and with decreasing slope; in the Oppenheimer-Snyder case there is an additional diverging factor $1/\text{sin}\,\tau$, i.e. the slope becomes more vertical more quickly, due to the geometry satisfying certain smoothness conditions at the origin (before the singularity forms).\par
This model serves as a good example of the behaviour of the inner horizon and its surface gravity in gravitational collapse which results in the formation of a Schwarzschild black hole. In more general scenarios of collapse the inner horizon may not reach the origin, instead halting a finite distance away. This is the case in the formation of a charged black hole, as we saw earlier, and also occurs when rotating and regular black holes form. Classically the dynamics of this horizon are restricted to either moving inward or halting, as moving back out would require the violation of causality, which in turn requires a matter source of negative energy density (i.e. violating the null energy condition) \cite{Hayward1993}. Semiclassically, however, there are no such restrictions, as we know well from Hawking evaporation.

\subsection{Inner horizon instability in classical and semiclassical gravity}

As we have seen above, there exist solutions for spacetimes with a static inner horizon (or with one which tends toward being so) which lasts for the entire lifetime of the outside universe. A general characteristic of these spacetimes is that they are incomplete (but extendable) for a set of geodesics which do not fall into any singularity within the bounds of the lifetime of the outside universe. The surface which separates these geodesics from their possible extensions is known as a Cauchy horizon, as it is the limit beyond which the evolution of spacetime can no longer be determined by initial conditions from a past (partial) Cauchy surface \cite{Wald1984,Hawking1973}.\par
However, as it turns out, this is an incomplete picture. The existence of a Cauchy horizon seems to rely on restrictive symmetry requirements used in idealised geometric constructions, and adding some perturbations (necessary to represent a generic physical scenario) results in the so-called mass-inflation instability \cite{PoissonIsrael89} (see \cite{Hamilton2008} for a review). Energy perturbations coming in from the outside universe and approaching the Cauchy horizon suffer a tremendous blueshift which tends to a divergence at the horizon itself. When back-reaction of the perturbations on the geometry is calculated, there seems to be a tendency towards the formation of a singularity which replaces (or overlaps with) this horizon. This singularity is either spacelike \cite{Hamilton2017}, or null \cite{Ori1991,Dafermos2017}. In the latter case, it also appears to be ``weak" \cite{Tipler1977,Ori1992}, in the sense that the spacetime is still extendable beyond it, though not smoothly, and there are no divergent tidal forces.\par
Geometries with a Cauchy horizon are quite easy to construct. All it takes is a metric of the type \eqref{metric1} with a redshift function $f(r)$ which has a zero with a negative slope. For simplicity, let us work in advanced Eddington-Finkelstein coordinates, in which the metric has the form
\begin{equation}\label{metric-vr}
ds^2=-f(r)dv^2+2dvdr+r^2d\Omega^2.
\end{equation}
We can expand the redshift function around its zero, which we say is at $r_{\rm i}$, as
\begin{equation}
f(r)\simeq-2k_1(r-r_{\rm i}),
\end{equation}
with $k_1>0$. Given the geodesic equation for radial trajectories $(v(\sigma),r(\sigma))$
\begin{equation}
\ddot{v}=-\frac{\partial_rf}{2}\dot{v}^2,
\end{equation}
where the dot indicates a derivative with respect to the affine parameter $\sigma$, with the linear order of $f(r)$ we get that the relation between the advanced time $v$ and the affine parameter $\sigma$ of geodesics which remain in the vicinity of $r_{\rm i}$ is
\begin{equation}\label{proptime}
\sigma\propto e^{-k_1v}.
\end{equation}
For outgoing null geodesics which tend toward the inner horizon asymptotically, this relation becomes exact.\par
For a spherically-symmetric geometry with an inner horizon to manifest the mass-inflation instability, there are three ingredients necessary, all of which are fairly generic:
\begin{enumerate}[label=(\roman*)]
	\item The first is the presence of an ingoing flux of energy from the outside which decays no quicker than the inverse of a polynomial in $v$, but quickly enough to be integrable. This, if nothing else, is a reasonable representation of the effects of infalling gravitational waves after the formation of a black hole, as was the original motivation in \cite{PoissonIsrael89}.
	\item The second is a condition on the response of the background to this flux. In particular, as the total infalling energy should be finite, there is a static geometry which is the asymptotic limit in time of the spacetime in question. The condition is for $r_{\rm i}$ to tend toward its asymptotic position slower than the exponential in \eqref{proptime}. This occurs e.g. in the Reissner-Nordström geometry \eqref{RN} for the inverse-polynomial flux.
	\item The third is an arbitrarily small outgoing null flux of energy inside the trapped region (which can represent e.g. the backscattering of the ingoing radiation). Following geodesic motion, the radial position of each ray within this flux tends toward the inner horizon from the outside exponentially quickly. Then, due to its proximity to the inner horizon, when the ingoing flux passes through it and displaces the position of the horizon, this outgoing flux acquires a large amount of energy from the change in gravitational potential, which is converted into mass. The region in the causal future of these two fluxes, i.e. part of the inside of the black hole, acquires a mass function which grows exponentially quickly in $v$, causing the mass inflation singularity as $v$ tends to infinity.
\end{enumerate}

Nevertheless, even with this instability, classical physics does not provide sufficient impediment to the presence of Cauchy horizons. This is due to the fact that it has been shown in several instances (e.g. \cite{Ori1991,Dafermos2017,Ori1998}) that the mass-inflation singularity can be weak, in the sense that there are no divergent tidal forces and the geometry can be continuously extended from there. Moreover, if the spacetime contains a cosmological constant, the singularity can be made to completely disappear in special cases in which the value of the surface gravity of the cosmological horizon is greater than that of the Cauchy horizon \cite{Mellor1992,Brady1992,Israel1991}. Thus, a complete argument against Cauchy horizons and their strange physical implications is to be looked for elsewhere, and indeed already exists.\par
As we know, considering classical matter as the source of gravity is just an effective description, as matter appears to be fundamentally quantum in nature. Although the interactions between quantum matter and spacetime are not fully understood, we have a good first approximation: semiclassical gravity (see e.g. \cite{BD}). This approach consists of defining quantum fields on a classical curved background spacetime and considering the backreaction from these fields on the spacetime geometry through the vacuum expectation value of the stress-energy operator constructed from a quantum field, renormalised in a manner which preserves covariance, i.e. the RSET defined above, which we will denote as $\expval{T_{\mu\nu}}^{\rm ren}$. This is usually considered as an addition to the right-hand side of the Einstein equations, where the term from the stress-energy tensor of matter which can be considered effectively classical is still maintained,
\begin{equation}
G_{\mu\nu}=8\pi G \left(T_{\mu\nu}^{\rm \,class}+\expval{T_{\mu\nu}}^{\rm ren}\right).
\end{equation}
The contribution from the RSET is, in most scenarios, negligible compared to that of classical matter and can be safely ignored. However, there are exceptions in which the RSET can become the dominant contribution, and one of them is spacetimes in which a stationary inner black-hole horizon is present, causing a Cauchy horizon to form \cite{Birrell1978,BalbinotPoisson93,Ori2019,Wald2019}.\par
With just a background given by the idealised symmetric and perturbation-free systems in which Cauchy horizons are present, there is an effect similar to mass inflation in the RSET, but which completely overshadows its classical counterpart in terms of magnitude. Calculating the RSET, or a sufficiently close approximation thereof, in such spacetimes gives an ingoing null component $\expval{T_{vv}}$ which has some non-zero value at the inner horizon, and switching to a time coordinate which is regular through the transition of the Cauchy horizon (e.g. the affine parameter in \eqref{proptime}) shows a physical divergence in this tensor. Backreaction from this effect alone is generally considered to lead to a ``strong" curvature singularity into which all incomplete geodesics fall \cite{Wald2019}, though due to the difficulty in calculating the RSET there is no complete semiclassically consistent solution to demonstrate this yet.\par
The difference in the degree of singularity between the classical and semiclassical approaches can be understood qualitatively by considering the stress-energy content of the exterior universe in each case. In the classical case there is an ingoing flux of energy falling into the black hole which decays sufficiently quickly so that the total energy injected into the system is finite. Although this does not make it obvious that the resulting singularity would be weak, it certainly comes as no surprise either.\par
On the other hand, taking the semiclassical approach, a black hole with an inner and outer horizon formed at finite time has an RSET which rather famously contains an outgoing flux of Hawking radiation on the outside, which does not trail off and if integrated throughout the lifetime of the outside universe is infinite. At the outer horizon itself there is a compensatory ingoing flux of negative energy \cite{DFU}, which again is constant and would again integrate to infinity. These infinities tell us that, if considered accumulatively, there is a large semiclassical source of dynamics in an otherwise classically static configuration. We must therefore take backreaction into account, even if the resulting dynamics appears negligible on small time scales.\par
Indeed, when we consider the RSET as an additional source of curvature we know that the outer horizon tends to move inward, evaporating the trapped region. Similarly, we find that the constant flux term at the inner horizon makes this horizon move outward, reducing the size of the trapped region from the other side. In fact, the source of the expected ``strong" singularity at the Cauchy horizon is nothing more than the result of initially ignoring the backreaction of this flux, the integration of which is then realised physically at the Cauchy horizon due to the growing discrepancy between $v$ and proper geodesic time.\par
The goal of this work is to correct this common oversight of only taking backreaction around the inner horizon into account when approaching the Cauchy horizon, where the RSET becomes large. Instead, we analyse the semiclassical dynamics of both horizons at a finite time form the point of view of the outside universe, and find as a result that an approach toward the formation of a Cauchy horizon is altogether unlikely, making the problem of whether there would be a singularity or an extension to another universe physically inconsequential.

\section{Backreaction on a black hole with an inner horizon}\label{static}

In this section we will present a toy model for the geometry of the formation of a regular black hole, which captures the main characteristics with sufficient generality, but is simple enough to allow analytical calculations with the semiclassical perturbations caused by the RSET. For the RSET we will use the Polyakov approximation \cite{Fabbri2005}. This approximation is obtained by integrating over the angular degrees of freedom, then quantising a field in the remaining 1+1 dimensions and calculating the RSET, and finally generalising the result to 3+1 dimensions. This approximate RSET captures the non-local effects at the outer horizon which lead to Hawking evaporation \cite{DFU,Fabbri2005}, as we show below, and we extend its use to the study of backreaction at the inner horizon.

\subsection{Geometric model and RSET}

The geometry we will be working with in this section is a simplified model of regular-black-hole formation. We start with a Minkowski spacetime, which has a line element
\begin{equation}
ds^2=-dv^2+2dvdr+r^2d\Omega^2
\end{equation}
in advanced Eddington-Finkelstein coordinates. After a point in time $v_{\rm f}$ the geometry becomes a regular, spherically symmetric static black hole,
\begin{equation}
ds^2=-f(r)dv^2+2dvdr+r^2d\Omega^2.
\end{equation}
The surface which separates the two regions can be seen as a collapsing null shell, a model often used when calculating semiclassical effects near the outer horizon (see e.g. \cite{SC2014}), though in this case it does not stem from a simple classical solution and is more of a geometric construct. The peeling of null geodesics, and consequently of modes of massless quantum fields, away from the outer horizon makes long-term semiclassical effects exhibit a certain universality there \cite{Barceloetal2010}. Thus a null shell is as good as any model of collapse if we want to study late-time Hawking evaporation. However, as we will see, at the inner horizon null geodesics behave in the opposite way to those at the outer one, i.e. they are accumulated, so the result is the exact opposite: late-time behaviour of semiclassical effects is highly sensitive to the characteristics of the collapse, and to conditions from the past of the spacetime in general. The collapsing shell model is therefore a simple geometric method for obtaining reasonable initial conditions for the quantum modes entering the black hole region, as doing so in a more generic collapse would require numerical computation. This method is also unique in the sense that it makes the quantum modes acquire the least amount of ``noise" from the collapse and serves to isolate the effects coming purely from the quantum field finding itself in the black hole geometry. Additionally, due to its simplicity it is an excellent example in which we can follow the origin and consequences of semiclassical effects by means of analytical expressions, as we will see.\par
\begin{figure}
	\centering
	\includegraphics[scale=.5]{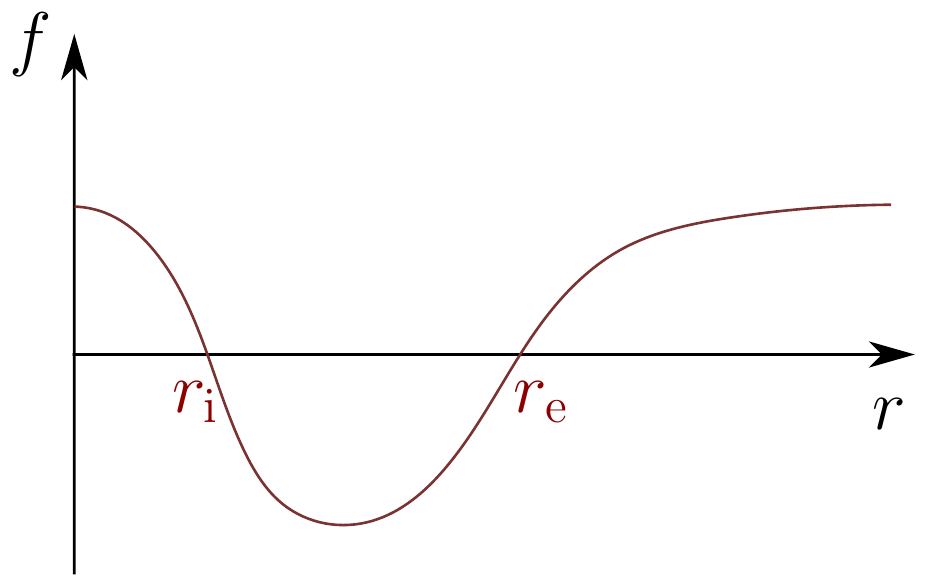}
	\caption{Redshift function $f(r)$ of our schematic regular black hole. For the inner horizon $r_{\rm h}=r_{\rm i}$ and for the outer one $r_{\rm h}=r_{\rm e}$.}
	\label{f1}
\end{figure}

For the redshift function in the black hole region $f(r)$ we will use a series expansion around each horizon of the form
\begin{equation}\label{redshift}
f(r)=2k_1(r-r_{\rm h})+2k_2(r-r_{\rm h})^2+2k_3(r-r_{\rm h})^3+\cdots,
\end{equation}
where $r_{\rm h}$ denotes the position of either the internal or external horizon and $k_{i}$ are constants, $k_1$ corresponding to the surface gravity of the horizon (negative for the internal and positive for the external horizon). Sufficiently close to each horizon, these two expansions are a valid representation of the redshift function, the global structure of which we assume is qualitatively of the form represented in fig. \ref{f1}. We note that throughout this work, when we construct a series assuming that a quantity with dimensions of length $l$ is ``small", we of course mean comparatively to the scale of the problem, i.e. that the sets of quantities $\{k_1l\}$, $\{k_2l^2,(k_1l)^2\}$, and subsequent orders, are progressively smaller.\par

\begin{figure}
	\centering
	\includegraphics[scale=.8]{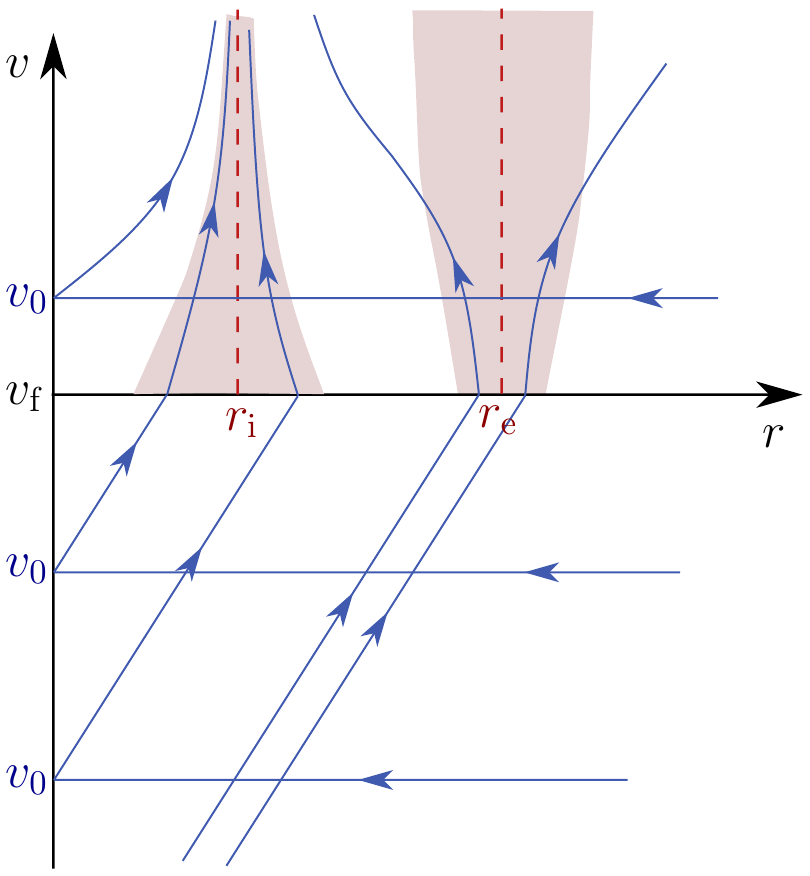}
	\caption{Representation of the trajectories of outgoing radial light rays. The black hole geometry begins at $v>v_{\rm f}$. The shaded regions around each horizon are a qualitative representation of the regions where our approximation for the RSET is valid.}
	\label{f2}
\end{figure}

We are thus treating an arbitrary black hole geometry of the form \eqref{metric-vr}. We use only one of the two degrees of freedom of the spherically symmetric spacetime for simplicity, given that the redshift function itself is enough to generate the causal structures we are interested in. The simple dynamical model for the formation of this structure, along with a focus on the areas around each horizon, will make analytical calculations of backreaction tractable. We can now proceed to construct the quantum ``in" vacuum state for this geometry, necessary for calculating the RSET. This vacuum state is defined from the Minkowski region at past null infinity, and its evolution is determined by the evolution of the spacetime. In particular, for the 1+1 dimensional approximation we will be using, its modes can be obtained from the behaviour of the radial null geodesics, represented in fig. \ref{f2}. The ingoing ones simply satisfy
\begin{equation}
v=\text{const.}
\end{equation}
For the outgoing ones we must solve the equation
\begin{equation}\label{outnull}
\frac{dr}{dv}=\frac{1}{2}f(r).
\end{equation}
For $v<v_{\rm f}$, $f(r)=1$ and the solution is simply
\begin{equation}\label{flatsol}
r(v)=\frac{1}{2}(v-v_0),
\end{equation}
where $v_0$ is an integration constant, identified as the time at which the light ray passes through the origin $r=0$. The ``in" vacuum is constructed from a pair of null coordinates $(v_{in},u_{in})$, which in this case are in fact $v_{\rm in}=v$ and $u_{\rm in}=v_0$ (i.e. in the $u_{\rm in}=\text{const.}$ outgoing null trajectories the value of $u_{\rm in}$ is the value of $v$ when said trajectories meet the origin). We have conveniently expressed the integration constant in terms of $v_0$, and we need to do the same with all solutions for outgoing light rays from here on, i.e. we need to trace them back to the origin. For the region in which the black hole has formed ($v>v_{\rm f}$), we are only interested in analysing the vicinity of each horizon, where \eqref{redshift} is sufficiently accurate. There, the solutions of \eqref{outnull} can be expressed in a series around $d_{\rm f}=0$, a parameter corresponding to the distance of the null ray from the horizon at $v_{\rm f}$,
\begin{equation}\label{df}
	d_{\rm f}=r(v_{\rm f})-r_{\rm h}=\frac{v_{\rm f}}{2}-\frac{v_0}{2}-r_{\rm h},
\end{equation}
the second equality coming from matching with \eqref{flatsol} at said time. For the purposes of this work we can express the solution up to order $d_{\rm f}^3$,
\begin{equation}\label{rv}
	\begin{split}
	r(v)\simeq r_{\rm h}&+e^{k_1(v-v_{\rm f})}d_{\rm f}+\frac{k_2}{k_1}\left[-e^{k_1(v-v_{\rm f})}+e^{2k_1(v-v_{\rm f})}\right]d_{\rm f}^2\\ &+\left[\left(\frac{k_2^2}{k_1^2}-\frac{k_3}{2k_1}\right)e^{k_1(v-v_{\rm f})}-2\frac{k_2^2}{k_1^2}e^{2k_1(v-v_{\rm f})}+\left(\frac{k_2^2}{k_1^2}+\frac{k_3}{2k_1}\right)e^{3k_1(v-v_{\rm f})}\right]d_{\rm f}^3.
	\end{split}
\end{equation}
At the outer horizon, where the surface gravity is positive, the coefficients of this series increase exponentially, making it a bad approximation for any finite $d_{\rm f}$ after sufficient time has passed. The reason for this can be seen in fig. \ref{f2}: outgoing light rays diverge away from the outer horizon, thus away form where the expansion \eqref{redshift} is valid. The reverse happens at the inner horizon, where the surface gravity is negative, as light rays converge toward this horizon. However, as we will see, this will translate inversely to the accuracy of the RSET approximation in each region.\par

To switch from $(v,r)$ to $(u_{in},v_{in})\equiv(u,v)$ coordinates, we use the relation
\begin{equation}
dr=\left.\frac{\partial r}{\partial v}\right|_{v_0=u}dv+\left.\frac{\partial r}{\partial v_0}\right|_{v_0=u}du,
\end{equation}
and note from \eqref{outnull} that $\partial r/\partial v|_{v_0=u}=f(r)/2$, which means that the term proportional to $dv^2$ cancels out. We are then left with the line element for the black hole region
\begin{equation}
ds^2=2\left.\frac{\partial r}{\partial v_0}\right|_{v_0=u}dudv+r^2(u,v)d\Omega^2=-C(u,v)dudv+r^2(u,v)d\Omega^2,
\end{equation}
where $C=-2\partial r/\partial v_0|_{v_0=u}$ is the conformal factor of the reduced 1+1 dimensional spacetime. From \eqref{rv} and \eqref{df} we obtain the expansion for this quantity up to order $d_{\rm f}^2$,
\begin{equation}\label{conformal}
	\begin{split}
	C\simeq\; &e^{k_1(v-v_{\rm f})}+2\frac{k_2}{k_1}\left[-e^{k_1(v-v_{\rm f})}+e^{2k_1(v-v_{\rm f})}\right]d_{\rm f}\\&+3\left[\left(\frac{k_2^2}{k_1^2}-\frac{k_3}{2k_1}\right)e^{k_1(v-v_{\rm f})}-2\frac{k_2^2}{k_1^2}e^{2k_1(v-v_{\rm f})}+\left(\frac{k_2^2}{k_1^2}+\frac{k_3}{2k_1}\right)e^{3k_1(v-v_{\rm f})}\right]d_{\rm f}^2.
	\end{split}
\end{equation}

With this we are ready to calculate the components of the RSET in the Polyakov approximation, given in ``in" coordinates by the standard expressions (see \cite{Fabbri2005})
\begin{subequations}\label{RSET}
	\begin{align}
	\expval{T_{uu}}&=\frac{1}{96\pi^2r^2}\left[\frac{\partial_u^2C}{C}-\frac{3}{2}\left(\frac{\partial_uC}{C}\right)^2\right],\\
	\expval{T_{vv}}&=\frac{1}{96\pi^2r^2}\left[\frac{\partial_v^2C}{C}-\frac{3}{2}\left(\frac{\partial_vC}{C}\right)^2\right],\\
	\expval{T_{uv}}&=\frac{1}{96\pi^2r^2}\left[\frac{\partial_uC\partial_vC}{C^2}-\frac{\partial_u\partial_vC}{C}\right].
	\end{align}
\end{subequations}
Calculating these quantities and switching back to $(v,r)$ coordinates we obtain the local expressions around each horizon
\begin{subequations}\label{RSETres}
	\begin{align}
		\expval{T_{vv}}&=-\frac{1}{96\pi^2r_{\rm h}^2}\frac{k_1^2}{2}+\order{r-r_{\rm h}},\\ \label{RSETrv}
		\expval{T_{rv}}&=-\frac{1}{96\pi^2r_{\rm h}^2}2k_2+\order{r-r_{\rm h}},\\
		\expval{T_{rr}}&=\frac{1}{96\pi^2r_{\rm h}^2}3\frac{k_3}{k_1}\left[1-e^{-2k_1(v-v_{\rm f})}\right]+\order{r-r_{\rm h}}.
	\end{align}
\end{subequations}
There are several things to note here. The first one is that we can actually obtain one additional order in the expansion of $\expval{T_{rv}}$, and two in the expansion of $\expval{T_{vv}}$, from just the terms in \eqref{conformal}, as only deriving with respect to $u$ reduces its order. This will be useful later in our calculation.\par
The second thing to note is the exponential in $\expval{T_{rr}}$. Such time-dependent coefficients are also present in higher-order terms in the remaining RSET components, and they affect the distance from $r_{\rm h}$ for which the truncated series expansions are a good approximation. In particular, due to the $C$ in the denominator of \eqref{RSET} and the subsequent change of coordinates, they will be exponentials of positive multiples of $(-k_1v)$. This means that for $k_1>0$ (outer horizon) the coefficients in the series quickly tend to constants, while for $k_1<0$ (inner horizon) they grow exponentially. This is the inverse effect on precision we were referring to earlier, which can be interpreted physically by observing fig. \ref{f2}. The light rays converge toward the inner horizon $r_{\rm i}$ so that determining the RSET in a region around it requires past information from larger and larger regions, where \eqref{redshift} is no longer valid. On the other hand, the diverging light rays away from the outer horizon imply that the approximation breaks down only when said rays are sufficiently far away from this horizon for \eqref{redshift} to cease being precise, and not due to incoming information. The latter case can be seen as a consequence of the universality of quantum effects around the external horizon of a black hole, analysed in \cite{Barceloetal2010}, while the former shows the reverse being true for the inner horizon, making its long-term semiclassical dynamics more difficult to pinpoint. Still, the fact that $\expval{T_{rr}}$ evaluated at the inner horizon itself grows exponentially makes it clear that semiclassical effects are not to be ignored there, as they can overcome their Planck-scale suppression very quickly.\par
The third thing to note is the negative ingoing flux given by $\expval{T_{vv}}$ at both horizons. This term is clearly non-local in curvature and is there solely due to the presence of each horizon, being determined by their respective surface gravities. At the outer horizon it is this negative flux which drives Hawking evaporation, compensating the positive thermal flux at infinity, as discussed originally in \cite{DFU}. At the inner horizon one may therefore expect that this term would lead to a similar non-causal behaviour of the spacetime, i.e. increasing the radial position of this horizon and evaporating the trapped region from the inside. This indeed seems to occur, as we show below.\par
Finally, it is worth making some remarks regarding the approximation which we use for the RSET \eqref{RSET}. One of its most obvious drawbacks is the divergence it generally has at $r=0$. If either horizon in our model were close to the origin (in terms of the Planck length), we would have to regularise this tensor (see e.g. \cite{Parentani1994,Arrechea2019}) or use a different approximation. However, we can simply restrict our geometric models to ones in which $r_{\rm h}\gg l_{\rm p}$, which, for black holes, is also a requirement for the semiclassical approximation itself to be valid.\footnote{This assumption is also valid for the dynamical cases studied in the next section, perhaps with the exception of the $n<1$ case of eq. \eqref{a1div} in its final stages, which however occurs only after the bouncing effect we are interested in, and well outside the range of validity of the approximation for the RSET that we use.}\par
As stated above, the reason why we use the Polyakov approximation is twofold. First, there is as yet no method to compute the exact 3+1 dimensional RSET for the ``in" vacuum with such generality; and second, this approximation seems to be enough to capture horizon-related effects, that is, at least when it comes to Hawking evaporation. However, it is easy to see that the terms local in curvature would differ from the exact RSET by just looking at the trace anomalies in 1+1 and 3+1 dimensions. The former is directly proportional to the $\expval{T_{rv}}$ term \eqref{RSETrv}, which depends only on the coefficient $k_2$ from \eqref{redshift}, while the latter, calculated with the expression given in e.g. \cite{BD}, is at zeroth order in $(r-r_{\rm h})$ a completely different function of the coefficients $k_1$, $k_2$ and $k_3$. We therefore do not exclude the possibility that an exact 3+1 dimensional calculation of backreaction may lead to different dynamics. In what follows it is shown, however, that the first perturbations on the position of either horizon are driven by the non-local flux term in $\expval{T_{vv}}$, making the resulting initial dynamics a robust result whenever such a flux is present.\par

\subsection{Perturbed Einstein equations}

In order to see the dynamical implications of backreaction near the two horizons, we will perturb the metric while maintaining spherical symmetry and equate the first order perturbation of the Einstein tensor to the RSET. Without loss of generality, the perturbed metric can be written as
\begin{equation}\label{metric-perturbed}
ds^2=-\left[f(r)+\delta f(v,r)\right]dv^2+2\left[1+\delta g(v,r)\right]dvdr+r^2d\Omega^2.
\end{equation}
Since we have expanded the RSET in powers of $(r-r_{\rm h})$, we must do the same with these perturbations:
\begin{equation}\label{dfg}
\begin{split}
\delta f(v,r)&=\delta f_0(v)+\delta f_1(v)(r-r_{\rm h})+\cdots,\\
\delta g(v,r)&=\delta g_0(v)+\delta g_1(v)(r-r_{\rm h})+\cdots.
\end{split}
\end{equation}
It is worth noting that since the RSET that we are using is fixed entirely by the background, the matter side of the Einstein equations may also require that a perturbation of the classical matter (the stress-energy tensor of which generates the background) be considered. However, with the series expansion which we are using, we have found that at the order needed to determine $\delta f_0$ and $\delta g_0$ the equations are compatible with this additional perturbation being zero.\par
We begin by equating the first order in $\delta f$ and $\delta g$ and zeroth order in $(r-r_{\rm h})$ of the Einstein tensor to the RSET \eqref{RSETres} times $8\pi l_{\rm p}^2$, where $l_{\rm p}$ is the Plank length. From the $vv$ component we obtain
\begin{equation}\label{vv}
(1-2k_1r_{\rm h})\delta f_0(v)-r_{\rm h}\delta f'_0(v)=-\frac{l_{\rm p}^2}{24\pi}k_1^2.
\end{equation}
Let us first analyse the implications of this equation in the case of the external horizon, with $r_{\rm h}=r_{\rm e}$ and $k_1>0$. If we take the familiar Schwarzschild case, where $2k_1r_{\rm e}=1$, with the initial condition $\delta f(v_{\rm f},r)=0$ the equation simply integrates to
\begin{equation}
\delta f_0=\frac{l_{\rm p}^2k_1^2}{24\pi r_{\rm e}}(v-v_{\rm f})=\frac{l_{\rm p}^2k_1^3}{12\pi}(v-v_{\rm f}).
\end{equation}
Therefore, the redshift function tends to increase around the horizon, leading to what can be identified as the first stages of Hawking evaporation of the trapped region. This can be seen explicitly by noting from the metric \eqref{metric-perturbed} that the modified equation for the trajectory of outgoing null radial geodesics is
\begin{equation}\label{outnullp}
\frac{dr}{dv}=\frac{1}{2}\frac{f(r)+\delta f(v,r)}{1+\delta g(v,r)}.
\end{equation}
Equating the right-hand side to zero and substituting the series \eqref{dfg}, we see that the first order change in the radial position of the external horizon is
\begin{equation}\label{linevap}
r_{\rm e}\quad\to\quad r_{\rm e}-\frac{\delta f_0}{2k_1}=r_{\rm e}-\frac{l_{\rm p}^2k_1^2}{24\pi}(v-v_{\rm f}).
\end{equation}
In other words, the Schwarzschild horizon has a tendency to shrink, albeit slowly.\par
For a non-Schwarzschild horizon (e.g. Reissner-Nordström, Schwarzschild-dS or -AdS, regular black hole models, etc.), the general solution of \eqref{vv} is
\begin{equation}\label{df0}
\delta f_0=\frac{l_{\rm p}^2k_1^2}{24\pi(2k_1r_{\rm e}-1)}\left[1-e^{-\frac{1}{r_{\rm e}}(2k_1r_{\rm e}-1)(v-v_{\rm f})}\right].
\end{equation}
In the limit $2k_1r_{\rm e}\to 1$ we recover the above Schwarzschild case. For other cases we can understand the initial tendencies by expanding this expression around $v=v_{\rm f}$,
\begin{equation}\label{dfexp}
\delta f_0=\frac{l_{\rm p}^2k_1^2}{24\pi r_{\rm e}}\left[(v-v_{\rm f})-\frac{2k_1r_{\rm e}-1}{2r_{\rm e}}(v-v_{\rm f})^2+\cdots\right].
\end{equation}
For a horizon with a surface gravity greater than that of a Schwarzschild black hole of the same size, i.e. for $k_1>1/(2r_{\rm e})$, we see that at linear order in $(v-v_{\rm f})$ the evaporation tends to be quicker while, when the $(v-v_{\rm f})^2$ term becomes important, it slows down. The opposite is true if $k_1<1/(2r_{\rm e})$, that is, the evaporation begins slower than in Schwarzschild but then tends to quicken. Of course, no definite conclusions can be drawn regarding the long-term evolution of the horizon due to the various approximations involved. Particularly, even if the above exponentials are good approximations initially, the fact that for an external horizon the coefficient $2k_1r_{\rm e}-1$ can change sign as the surface gravity and radius evolve makes it likely that the overall evolution has a kind of intermediate behaviour, more akin to the Schwarzschild case.\par
To complete our picture of what occurs at the outer horizon, we can look at the rest of the perturbed Einstein equations in search for a solution for $\delta g_0$. Combining the $vr$ component at zero order in $(r-r_{\rm e})$ and the $vv$ component at first order [as we mentioned above, the first and second orders of $\expval{T_{vv}}$ can be obtained easily from \eqref{conformal}], we get
\begin{equation}
\begin{split}
(2k_1r_{\rm e}-1)\delta g'_0(v)+\frac{1}{r_{\rm e}}&(1-4k_1r_{\rm e})\delta g_0(v)=\\&\frac{1}{r_{\rm e}}(1+2k_1r_{\rm e}+4k_2r_{\rm e}^2)\delta f_0(v)+\frac{l_{\rm p}^2}{6\pi r_{\rm e}}k_2(1-2k_1r_{\rm e})
\end{split}
\end{equation}
In the Schwarzschild case, where $k_1=1/(2r_{\rm e})$ and $k_2=-1/(2r_{\rm e}^2)$, this reduces to
\begin{equation}
\delta g_0=0.
\end{equation}
In more general scenarios, with different relations between $r_{\rm e}$ and the coefficients $k_i$, $\delta g_0$ is a non vanishing function of time which, looking at \eqref{outnullp}, causes a slightly faster or slower (depending on its sign) divergence of null geodesics away from the horizon. From the metric \eqref{metric-perturbed} its effect can also be interpreted physically as a contraction or expansion of space in the radial direction.\par
Having recovered the familiar evaporative tendency of the outer horizon, we can now move on to the analysis of backreaction at the inner horizon. The equations are exactly the same, except that for the sake of notation we switch $r_{\rm e}\to r_{\rm i}$ and we have to keep in mind that $k_1$ is now negative. To see how the position of the inner horizon changes we must look at \eqref{df0}, which we can rewrite in a more convenient manner given the sign of $k_1$ as
\begin{equation}\label{df0i}
\delta f_0=\frac{l_{\rm p}^2k_1^2}{24\pi(1-2k_1r_{\rm i})}\left[e^{\frac{1}{r_{\rm i}}(1-2k_1r_{\rm i})(v-v_{\rm f})}-1\right].
\end{equation}
We see that, much like before, the redshift function tends to increase, and this leads to a reduction of the size of the trapped region,
\begin{equation}
r_{\rm i}\to r_{\rm i}+\frac{\delta f_0}{2|k_1|}.
\end{equation}
The series expansion around $v=v_{\rm f}$ is the same as \eqref{dfexp}, meaning that if the absolute value of the surface gravity of the inner horizon is greater than that of the outer horizon (which is usually the case, except in near-extremal configurations), the initial tendency is for the trapped region to begin evaporating more quickly from the inside than from the outside. Additionally, we note that the coefficient multiplying $v$ in the exponential in $\delta f_0$ is positive for any inner horizon, implying the possibility that the exponential behaviour for the evaporation may be a more general characteristic which is maintained beyond the regime of validity of our approximation. We will analyse this more closely for a particular family of dynamical solutions in the following.

\section{Dynamical horizons and self-consistent solutions}\label{dynamic}

We have seen how the perturbations on the metric caused by the RSET behave around the static inner and outer horizons of a black hole. But either due to dynamics in the classical sector or to backreaction itself, the position of these horizons is generally not static. Calculating the RSET on a dynamical background can be challenging even in the Polyakov approximation, but we will work around this difficulty in a manner similar to what we employed in the previous section. We will expand the redshift function $f(v,r)$ in a series around a dynamical horizon $r_{\rm h}(v)$ and calculate the RSET in terms of the time-dependent coefficients of this series. This will allow us to again obtain the deviations in the metric $\delta f(v,r)$ and $\delta g(v,r)$ around each horizon. Doing so without completely specifying the dynamics of the background will then allow us to arrive at approximate self-consistent solutions in some particularly simple cases.\par

\subsection{RSET around dynamical horizons}

To maintain the simplicity of the initial conditions for the quantum modes we had in the previous section, we will once again consider a spacetime model which is flat up to a time $v_{\rm f}$ and then transitions abruptly to a black hole geometry. The relation between the parameter $d_{\rm f}$ and the ``in" coordinate $u$, identified with $v_0$, is therefore still given by \eqref{df}, with $r_{\rm h}$ in this case being $r_{\rm h}(v_{\rm f})$.\par
The black hole geometries which we will use for a background are of the form
\begin{equation}
ds^2=-f(v,r)dv^2+2dvdr+r^2d\Omega^2.
\end{equation}
We note that considering both degrees of freedom of the spherically-symmetric geometry yields an equally straightforward calculation for the RSET, the main complication arising at the stage of resolution of the perturbed equations. We will use the above form because the simplified cases in which we will analyse backreaction maintain it. The expansion we will use for the redshift function is completely analogous to the one in the static case,
\begin{equation}
f(v,r)=2k_1(v)[r-r_{\rm h}(v)]+2k_2(v)[r-r_{\rm h}(v)]^2+2k_3(v)[r-r_{\rm h}(v)]^3+\cdots.
\end{equation}
Assuming that the quantity $r-r_{\rm h}(v)$ is small, the trajectories of outgoing null radial geodesics can be expanded as
\begin{equation}
r(v)=r_{\rm h}(v)+p_1(v)+d_{\rm f}e^{\tilde{k}_1(v)}+p_2(v,d_{\rm f})+p_3(v,d_{\rm f})+\cdots,
\end{equation}
where
\begin{align}
\tilde{k}_1(v)&=\int_{v_{\rm f}}^vk_1(\tilde{v})d\tilde{v},\\
p_1(v)&=-e^{\tilde{k}_1(v)}\int_{v_{\rm f}}^ve^{-\tilde{k}_1(\tilde{v})}r'_{\rm h}(\tilde{v})d\tilde{v}, \label{p1}\\
p_2(v,d_{\rm f})&=e^{\tilde{k}_1(v)}\int_{v_{\rm f}}^ve^{-\tilde{k}_1(\tilde{v})}k_2(\tilde{v})\left[p_1(\tilde{v})+d_{\rm f}e^{\tilde{k}_1(\tilde{v})}\right]^2d\tilde{v},\\
\begin{split}
p_3(v,d_{\rm f})&=e^{\tilde{k}_1(v)}\int_{v_{\rm f}}^ve^{-\tilde{k}_1(\tilde{v})}\left\{2k_2(\tilde{v})p_2(\tilde{v})\left[p_1(\tilde{v})+d_{\rm f}e^{\tilde{k}_1(\tilde{v})}\right]\right.\\&\hspace{3.6cm}\left.+k_3(\tilde{v})\left[p_1(\tilde{v})+d_{\rm f}e^{\tilde{k}_1(\tilde{v})}\right]^3\right\}d\tilde{v},
\end{split}
\end{align}
with $r'_{\rm h}(v)=dr_{\rm h}/dv$. Unlike in the static case, this expansion is performed around the first order solution for the separation from the horizon, $r_1(v)\equiv\left[p_1(\tilde{v})+d_{\rm f}e^{\tilde{k}_1(\tilde{v})}\right]$ (in units of the scale of each horizon), making it progressively worse with time no matter how small the initial parameter $d_{\rm f}$ is. In particular, looking at the expression for $p_1$, the larger the rate of change of the horizon position $r'_{\rm h}(v)$ is, the quicker the approximation breaks down. However, this can be delayed if the coefficients $k_i(v)$ with $i\ge 2$ remain small enough compared to powers of $k_1(v)$, as we will consider in one of our simplified models below.\par
With these expressions we can obtain a generalisation of \eqref{RSETres} for dynamical backgrounds, valid for a small but finite time interval. This RSET at zero order in $r_1(v)$ is
\begin{subequations}\label{RSETdyn}
\begin{align}
\expval{T_{vv}}&\simeq-\frac{1}{96\pi^2r_{\rm h}(v)^2}\left[\frac{1}{2}k_1(v)^2-k'_1(v)\right],\label{Tvv} \\
\expval{T_{rv}}&\simeq-\frac{1}{96\pi^2r_{\rm h}(v)^2}2k_2(v),\\
\begin{split}
\expval{T_{rr}}&\simeq\frac{1}{96\pi^2r_{\rm h}(v)^2}e^{-2\tilde{k}_1(v)}\left\{\int_{v_{\rm f}}^v\left[8k_2(\tilde{v})e^{\tilde{k}_1(\tilde{v})}\int_{v_{\rm f}}^{\tilde{v}}k_2(\bar{v})e^{\tilde{k}_1(\bar{v})}d\bar{v}+6k_3(\tilde{v})e^{2\tilde{k}_1(\tilde{v})}\right]d\tilde{v}\right.\\ &\hspace{4cm}\left. -6\left[\int_{v_{\rm f}}^vk_2(\tilde{v})e^{\tilde{k}_1(\tilde{v})}d\tilde{v}\right]^2\right\},
\end{split}
\end{align}
\end{subequations}
from which we can easily recover \eqref{RSETres} in the static limit (paying attention to the integration limits). As a side note, it is interesting to see that the expression for $\expval{T_{vv}}$ has the same structure as the one found in \cite{Barbado2016} for the difference between the values of the RSET in two different vacuum states, when expressed in terms of the effective temperature function.

\subsection{Perturbed equations and Hawking evaporation}

With an expansion analogous to \eqref{dfg} for the metric perturbations, the generalisation of eq. \eqref{vv} is now
\begin{equation}
\left[1-2r_{\rm h}(v)k_1(v)\right]\delta f_0(v)-r_{\rm h}(v)(\delta f')_0(v)-2k_1r_{\rm h}(v)r'_{\rm h}(v)\delta g_0(v)=-\frac{l_{\rm p}^2}{24\pi}\left[k_1^2(v)-2k'_1(v)\right],
\end{equation}
where $(\delta f')_0=\delta f'_0-\delta f_1r'_{\rm h}$ is the zeroth order of the derivative of $\delta f$ with respect to $v$ (while $\delta f'_0$ is the derivative of the zeroth order; the two only coincide for static backgrounds). We see that the simple decoupling we had for $\delta f_0$ in the static case is not present in general, unless the rate of change of the background is small enough for $r'_{\rm h}(v)\ll k_1(v)r_{\rm h}(v)$ to be satisfied. If the dynamics are induced only by backreaction itself, then $r'_{\rm h}$ and $k'_1$ are initially of order $l_{\rm p}^2$ (in dimensions of the horizon scale), and so are $\delta f$ and $\delta g$, implying that the approximation
\begin{equation}\label{vv-dyn}
\left[1-2r_{\rm h}(v)k_1(v)\right]\delta f_0(v)-r_{\rm h}(v)\delta f'_0(v)\simeq -\frac{l_{\rm p}^2}{24\pi}k_1^2(v),
\end{equation}
is reasonable. $r'_{\rm h}$ being small also implies that the series expansion itself is accurate for a longer period of time, as can be seem from \eqref{p1}.\par
Let us once again start by looking at the evolution of an initially static external Schwarzschild horizon. We have the initial condition $2r_{\rm e}(v_{\rm f})k_1(v_{\rm f})=1$, and possible deviations from this equation at later times $v$, once multiplied by $\delta f_0$, are of the same order as the terms we have neglected above, allowing us to now neglect the first term. We also have the relation $\delta f'_0 \simeq -r'_{\rm e}/r_{\rm e}$ for the displacement of the Schwarzschild horizon. Substituting into this equation we get
\begin{equation}\label{hawk}
r'_{\rm e}\simeq-\frac{l_{\rm p}^2}{96\pi}\frac{1}{r_{\rm e}^2}.
\end{equation}
The solution of this approximate self-consistent equation is simply
\begin{equation}\label{schw}
r_{\rm e}(v)\simeq \left[r_{\rm e}(v_{\rm f})^3-\frac{l_{\rm p}^2}{32\pi}(v-v_{\rm f})\right].
\end{equation}
To check that we are on the right track, we can expand this solution around $v_{\rm f}$ and see that we recover the linear tendency form \eqref{linevap}. Additionally, we can extrapolate this solution to later times, assuming that \eqref{schw} remains approximately valid at the later stages of evaporation, and get the Hawking evaporation time
\begin{equation}\label{hawk-evap}
v_{\rm H}\simeq 256\pi \frac{M^3}{l_{\rm p}^2}\simeq \left(\frac{M}{M_\odot}\right)^3 10^{73}\,\text{s},
\end{equation}
where $M=r_{\rm e}(v_{\rm f})/2$ is the initial black hole mass and $M_\odot$ is the solar mass. The approximations involved in obtaining this extrapolation are equivalent to the quasi-stationary approximation used originally by Hawking to estimate the lifetime of black holes \cite{Hawking1974}.\par
In other scenarios where the quasi-stationary assumption is valid for a period of time, we can still use eq. \eqref{vv-dyn} to get a first glimpse at self-consistent solutions. We can integrate for $\delta f_0$, obtaining
\begin{equation}\label{df0dyn}
\delta f_0(v)=e^{\int_{v_{\rm f}}^v[1/r_{\rm h}(\tilde{v})-2k_1(\tilde{v})]d\tilde{v}}\int_{v_{\rm f}}^ve^{-\int_{v_{\rm f}}^{\tilde{v}}[1/r_{\rm h}(\bar{v})-2k_1(\bar{v})]d\bar{v}}\frac{l_{\rm p}^2k_1(\tilde{v})^2}{24\pi r_{\rm h}(\tilde{v})}d\tilde{v}.
\end{equation}
Taking the right-hand side as a function of a slowly-evolving classical background would just give a generalisation of \eqref{dfexp}. Modifying the right-hand side with the backreaction due to $\delta f$ and $\delta g$ would give a more accurate expression of backreaction in a quasi-stationary approximation, the validity of which would have to be checked along the evolution in each case. It is interesting to note that for an inner horizon, where $k_1(v)<0$, unless $k_1(v)$ quickly tends to zero, we once again have the growing exponential we had in \eqref{df0i}, now somewhat hidden in the term given by the lower bound of the integral which is outside the exponentials. To see how this exponential behaviour manifests itself in an approximate self-consistent solution, we will now look at a specific set of backgrounds for which we can solve this equation.\par

\subsection{Inner horizon evaporation for a simple background}

Let us consider for the black hole region a redshift function which around the inner horizon has the form
\begin{equation}\label{bg}
f(v,r)=1-\frac{\lambda(v)}{2}r-\frac{\alpha(v)}{r},
\end{equation}
with $\lambda(v)$ a positive function and $\alpha(v)$ a function which satisfies the initial condition $\alpha(v_{\rm f})=0$. While $2\lambda\alpha<1$, the position and surface gravity of the inner horizon within this geometry are
\begin{align}
r_{\rm i}&=\frac{1}{\lambda}(1+\sqrt{1-2\lambda\alpha})=\frac{2}{\lambda}-\alpha-\frac{1}{2}\alpha^2\lambda+\cdots, \label{ri}\\
k_1&=\frac{1-2\lambda\alpha-\sqrt{1-2\lambda\alpha}}{4\alpha}=-\frac{1}{4}\lambda+\frac{1}{8}\alpha\lambda^2+\cdots, \label{k1dyn}
\end{align}
where the series expansions on the right-hand side are valid while $|\alpha\lambda|\ll 1$.\par
For $\alpha=0$ the only non-zero term of the RSET on this background would be $\expval{T_{vv}}$ from \eqref{RSETdyn}, and the local approximation for null geodesics involved in obtaining it would actually be exact (i.e. higher order terms in the expansion would be zero) in a finite region around the inner horizon, akin to the left shaded region in fig. \ref{f2}. There, we would thus have
\begin{equation}\label{Tvvin}
\expval{T_{vv}}=-\frac{1}{96\pi^2r^2}\left[\frac{1}{2}k_1(v)^2-k'_1(v)\right].
\end{equation}
All this remains approximately true while $\alpha$ is small compared to $\lambda$ in units of $r_{\rm i}$, and we will use this fact to simplify the dynamical perturbation equations in order to obtain an approximate self-consistent solution.\par
The Einstein tensor of this geometry is
\begin{align}
G_{vv}&=\frac{\lambda(v)}{r}-\lambda(v)^2+\frac{\lambda'(v)}{2}-\lambda(v)\frac{\alpha(v)}{r^2}+\frac{\alpha'(v)}{r^2},\\
G_{vr}&=-\frac{\lambda(v)}{r},\\
G_{\theta\theta}&=\frac{G_{\varphi\varphi}}{{\rm sin}^2\theta}=-\frac{r\lambda(v)}{2}.
\end{align}
We see that $\alpha(v)$ appears only in two terms of the $vv$ component and is divided by $r^2$. We can consider these two terms as the ones sourced by the RSET \eqref{Tvvin}, the rest corresponding to the classical background, which fixes the function $\lambda(v)$ (keep in mind that the classical background is just a toy model used to construct the causal structure we are interested in). Then $\alpha(v)$ becomes the semiclassical perturbation satisfying the equation
\begin{equation}\label{alpha}
\alpha'(v)-\lambda(v)\alpha(v)=-\frac{l_{\rm p}^2}{24\pi}\left[k_1(v)^2-2k'_1(v)\right].
\end{equation}
Integrating for $\alpha$ we obtain
\begin{equation}\label{alphadyn}
\alpha(v)=-e^{\int_{v_{\rm f}}^v\lambda(\tilde{v})d\tilde{v}}\int_{v_{\rm f}}^ve^{-\int_{v_{\rm f}}^{\tilde{v}}\lambda(\bar{v})d\bar{v}}\frac{l_{\rm p}^2}{24\pi}\left[k_1(\tilde{v})^2-2k'_1(\tilde{v})\right]d\tilde{v}.
\end{equation}
If we take a background with $\lambda=\text{const.}$ and substitute $k_1$ for its zero-order value from the expansion \eqref{k1dyn}, the integration yields
\begin{equation}\label{alphastat}
\alpha(v)=-\frac{l_{\rm p}^2\lambda}{192\pi}\left[e^{\lambda(v-v_{\rm f})}-1\right].
\end{equation}
This solution is valid until the initially zero $\alpha$ becomes comparable to $\lambda$ (in units of $r_{\rm i}$), which, as \eqref{df0i} and \eqref{df0dyn} suggested, does not take long due to the growing exponential. Introduced into the geometry, this term behaves like a negative mass, which tends to move the inner horizon outward, as can be seen from \eqref{ri}. The exponential growth of this mass and the displacement of the inner horizon can be seen as a semiclassical manifestation of the inner horizon instability, with an opposite tendency to its classical counterpart.\par
Given this intriguing tendency to evaporate the trapped region from the inside, it is only natural to ask oneself what may happen if the same behaviour were to continue throughout the evolution of the inner horizon, up until the disappearance of the trapped region. In other words, what would the result be if the driving force of evaporation continued to be the local $\expval{T_{vv}}$ term on the right-hand side of \eqref{alpha}. Although this assumption is less justifiable dynamically than the analogous one used for Hawking evaporation in eq. \eqref{hawk}, one may think of it as just an extrapolation from the initial tendency. If nothing else, it serves as an example of how the dynamics of the inner horizon can continue with an RSET which continues violating the energy positivity conditions, as it seems likely to do around a horizon, making the geometry evolve in a non-causal manner.\par
To answer this question, we can take into account the change in surface gravity due to the evolution of $\alpha$ on the right-hand side of \eqref{alpha} through \eqref{k1dyn}. Writing $\alpha$ in terms of $k_1$ as
\begin{equation}
\alpha=\frac{1}{2}\frac{4k_1+\lambda}{(2k_1+\lambda)^2},
\end{equation}
equation \eqref{alpha} becomes
\begin{equation}\label{sc}
\left[\frac{4k_1}{(\lambda+2k_1)^3}+\frac{l_{\rm p}^2}{12\pi}\right]k'_1+\frac{\lambda'}{2}\frac{\lambda+6k_1}{(\lambda+2k_1)^3}+\frac{\lambda}{2}\frac{\lambda+4k_1}{(\lambda+2k_1)^2}-\frac{l_{\rm p}^2}{24\pi}k_1^2=0.
\end{equation}
This equation contains as solutions the initial behaviours in the backreaction problem given by \eqref{alphadyn} and \eqref{alphastat}, along with their extensions. More generally, it governs the evolution of a geometry whose dynamics is modified by a (usually negative) ingoing flux of energy determined by the surface gravity at its inner apparent horizon through \eqref{Tvvin}.\par
\begin{figure}
	\centering
	\includegraphics[scale=.45]{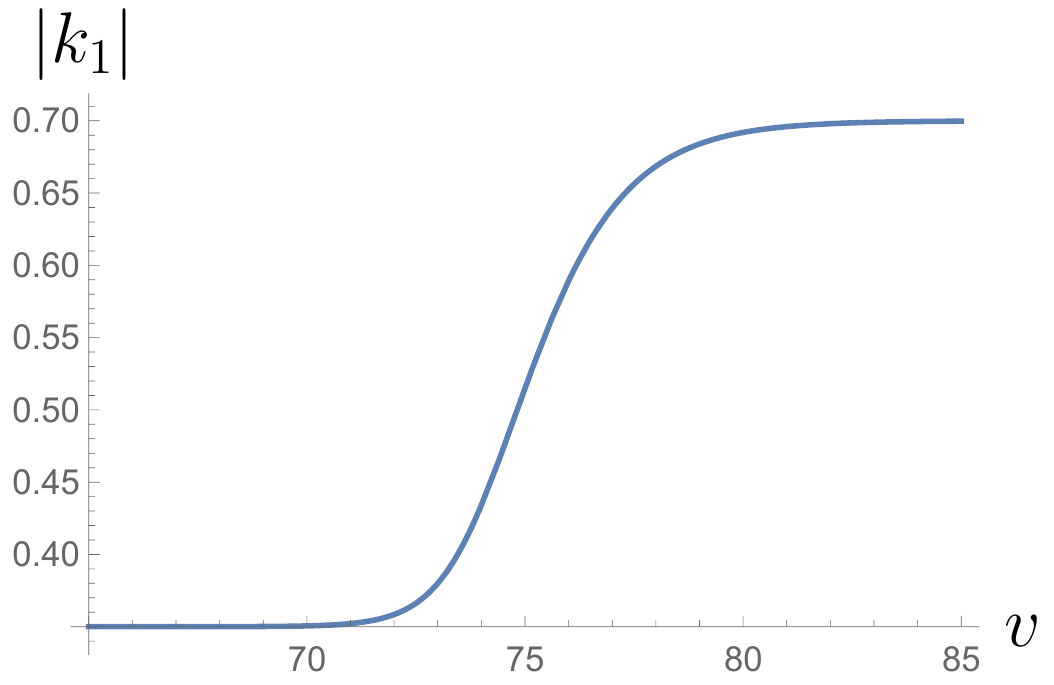}
	\includegraphics[scale=.45]{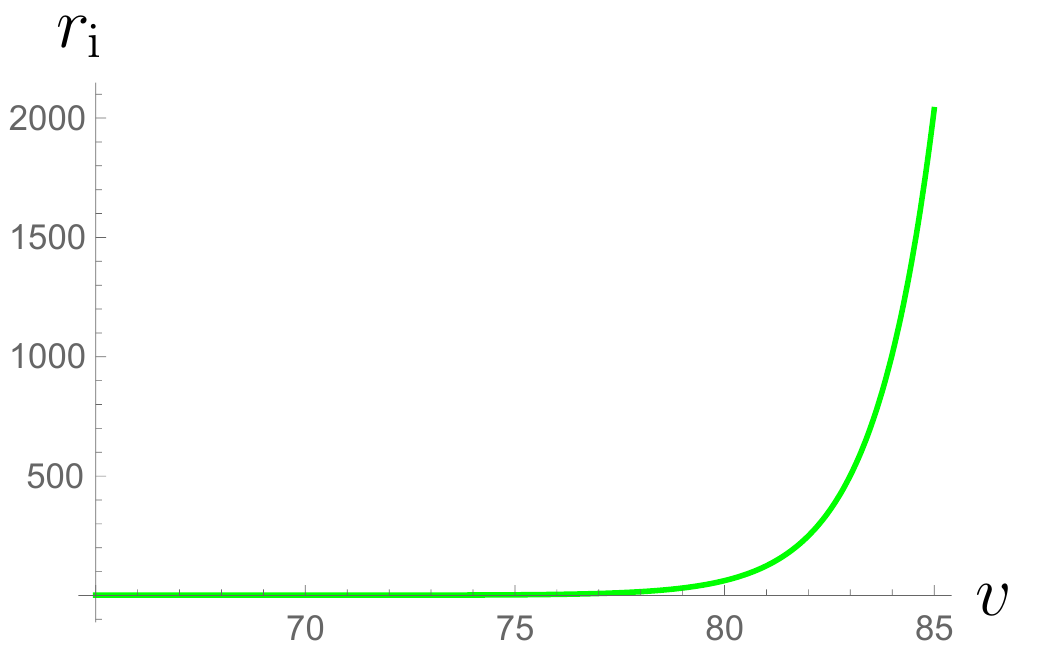}
	\includegraphics[scale=.45]{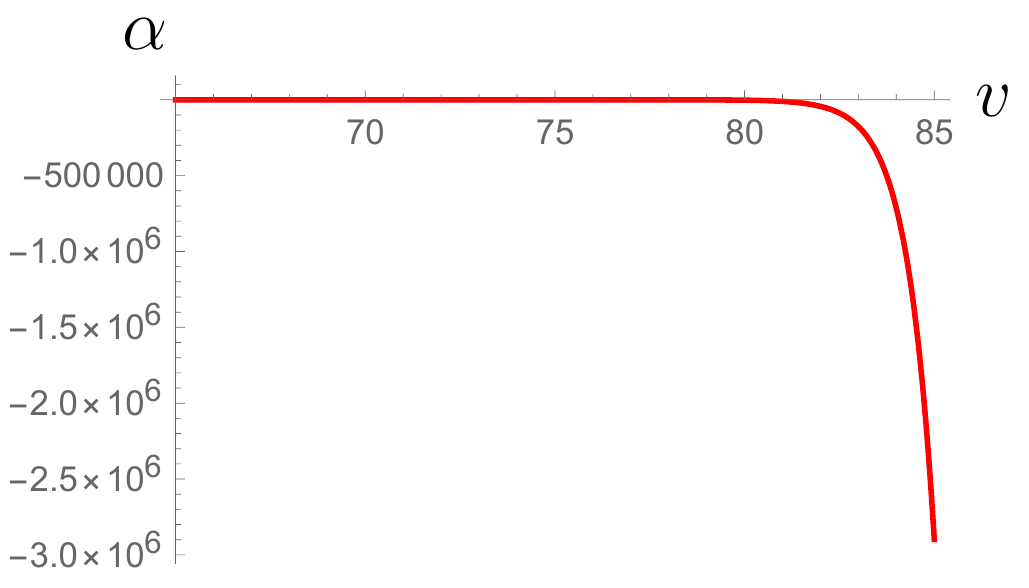}
	\caption{Plots of $|k_1(v)|$, $r_{\rm i}(v)$ and $\alpha(v)$, from left to right, for the evaporation of the inner horizon. We have taken $\lambda=0.7$ and $l_{\rm p}=10^{-8}$ (as a large difference in scales is required but smaller values of $l_{\rm p}$ make numerical evaluation more difficult and lead to no qualitative changes).}
	\label{inner-evap}
\end{figure}

Taking $\lambda$ as a positive constant, all solutions of \eqref{sc} have the same behaviour, shown in fig. \ref{inner-evap}: the decrease in $\alpha$ observed perturbatively initially increases the absolute value of $k_1$ but then makes it tend to a constant (with a value $\lambda/2$). This surface gravity then continues to feed the right-hand side of \eqref{alpha}, extending the exponential behaviour of $\alpha$ indefinitely. The radial position of the inner horizon also continues to increase exponentially.\par

In summary, the geometry \eqref{bg} fed by the flux \eqref{Tvvin} has an inner horizon which moves outward exponentially quickly. This goes on indefinitely due to the global structure of \eqref{bg}, which does not contain an outer bound for the trapped region. In a more realistic scenario, even if backreaction continues to be governed by a term like \eqref{Tvvin}, we expect such dynamics to end when the trapped region disappears. The main conclusion we can extract from this is that incorporating the modifications of the geometry due to backreaction on the right-hand side of \eqref{alpha} does not tend to decrease the rate of the initially exponential evaporation of the inner horizon.\par
If we assume that such a behaviour is the dominant factor in the elimination of a trapped region, we can estimate a revised evaporation time for black holes with an inner horizon. Considering a black hole of mass $M$, with an outer horizon $r_{\rm e}\sim M$ which evaporates slowly \textit{à la} Hawking, and an inner horizon with an initial position $r_{\rm i,0}$ and surface gravity $k_{1,0}$, the inner horizon would meet the outer one after a time
\begin{equation}
v_{\rm evap}\simeq \frac{1}{k_{1,0}+(2r_{\rm i,0})^{-1}}\log\frac{M}{l_{\rm p}}\lesssim\frac{M}{M_\odot}\times 10^{-5}\,\text{s},
\end{equation}
where $M_\odot$ is the solar mass, and we have obtained the upper bound on the right-hand side by assuming that the surface gravity at the inner horizon is initially greater than that of the outer horizon, the latter of which we take to be of the order $1/M$ (the logarithmic dependence has been omitted in this bound as for no astrophysically reasonable object would it increase the order of magnitude further). Needless to say, this process is much quicker than the time it would take for a Schwarzschild black hole to evaporate from the outside, given by \eqref{hawk-evap}.\par
Therefore, even if we are using the word ``evaporation" to describe the leading effects of semiclassical backreaction on the inner horizon, the rapidity of its outwards displacement brings to mind a more abrupt phenomenon that may be better described as an ``explosion", although this word should be understood with a different meaning than the one intended by Hawking.

\subsection{Collapsing matter: singularity or bounce}

So far our results in this section have been a direct generalisation of the perturbation analysis in the previous one. But the treatment on dynamical backgrounds and self-consistent extrapolation allow for a wider range of solutions to be analysed, in particular ones in which the classical backgrounds itself is dynamical. We expect the backreaction of a moving inner horizon to have a similar effect as observed for the initially static background: to push it outward and try to diminish the size of the trapped region. Whether and when this tendency from backreaction can overcome and dominate over the dynamics of the classical background is what we will analyse here.\par
What we will look at is the backreaction problem around the dynamical inner horizon of a gravitational collapse which would classically end in a Schwarzschild-like black hole. We construct a geometry around this horizon of the type \eqref{bg} with $\alpha=0$ (classically) and
\begin{equation}\label{a1div}
\lambda=\frac{\lambda_0^{1-n}}{(v_{\rm s}-v)^n},
\end{equation}
with $v\in(0,v_{\rm s})$, $n>0$ and $\lambda_0$ a constant (with the same dimensions as $\lambda$) which defines the characteristic length scale of the problem. Matching this with a Minkowski region through an ingoing null shell at $v=0$, we get a picture of a collapse in which the inner horizon initially travelled inward at light-speed but then slowed down before continuing to the centre. This is once again a method of simplifying the initial conditions for the quantum modes which enter the black hole region by removing their dependence on the details of the collapse in the far past, thus focusing only on the effects caused by these modes entering the vicinity of the inner horizon at the final stages of the collapse. This also goes hand in hand with our approximation \eqref{alpha}.\par
Introducing these backgrounds into eq. \eqref{alphastat}, we can analyse whether horizon-related semiclassical effects can become relevant to the overall dynamics. What we find is that there are two ways $\alpha$ can become large enough for this to occur. First, if the integral of $\lambda$ diverges, which is the case for $n\ge 1$, then $\alpha$ always diverges as the exponential of this integral, making it clearly dominant over the classical background. Second, regardless of whether the integral of $\lambda$ diverges or not, if the interval $(0,v_{\rm s})$ is large enough, i.e. if the background dynamics is slow enough for a long period of time, then an effect similar to what occurred with a static background may dominate. Then the exponential of the integral of $\lambda$ becomes large enough to overcome the Planck scale suppression, even though it may not tend to a divergence.\par
\begin{figure}
	\centering
	\includegraphics[scale=.5]{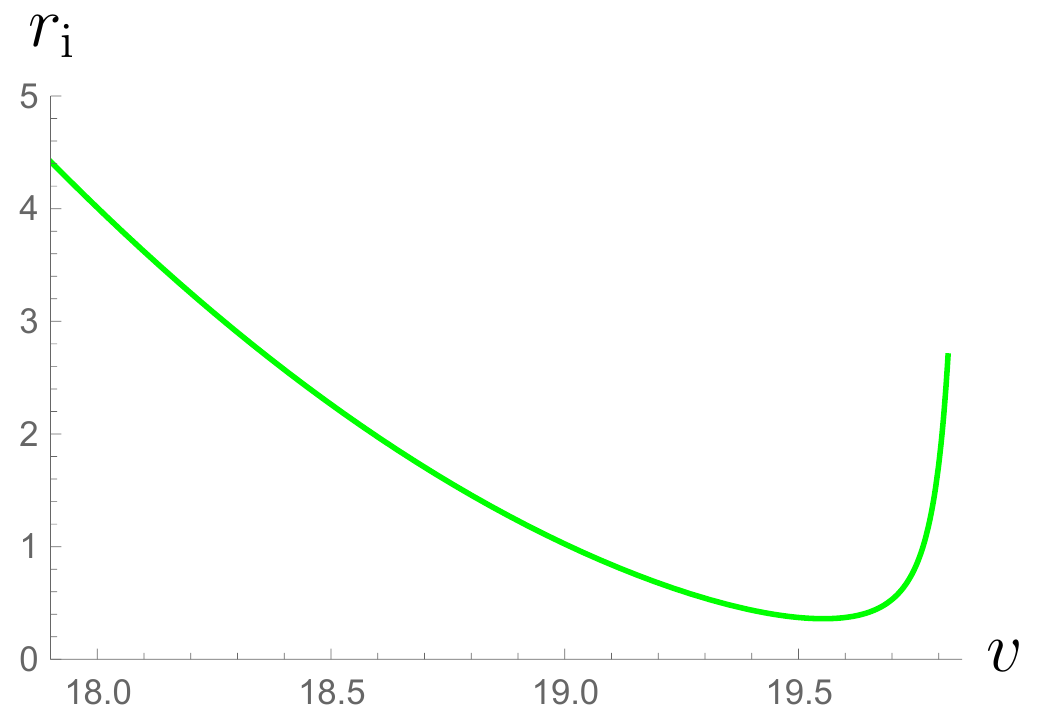}
	\includegraphics[scale=.5]{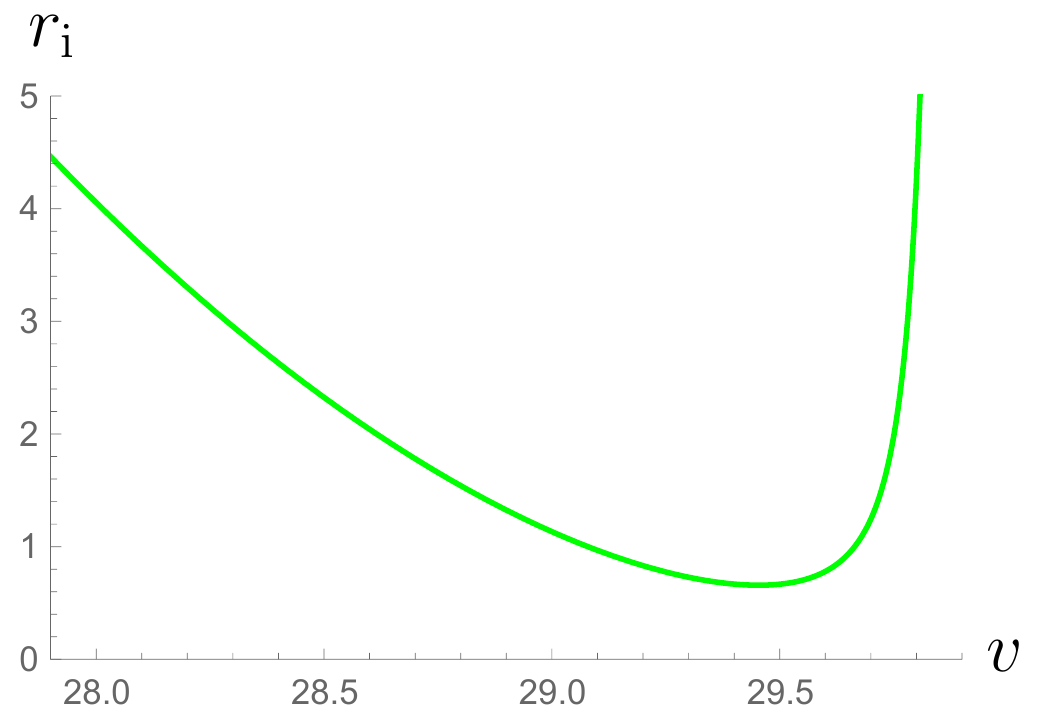}
	\caption{Plots of $r_{\rm i}(v)$ for a background given by \eqref{a1div} with $\lambda_0=1$ and $n=2$, with $v_{\rm s}=20$ on the left and with $v_{\rm s}=30$ on the right (starting from zero perturbation at $v=0$). Units are once again given by $l_{\rm p}=10^{-8}$, for the same computational and qualitative reasons as above.}
	\label{bounce2}
\end{figure}
\begin{figure}
	\centering
	\includegraphics[scale=.5]{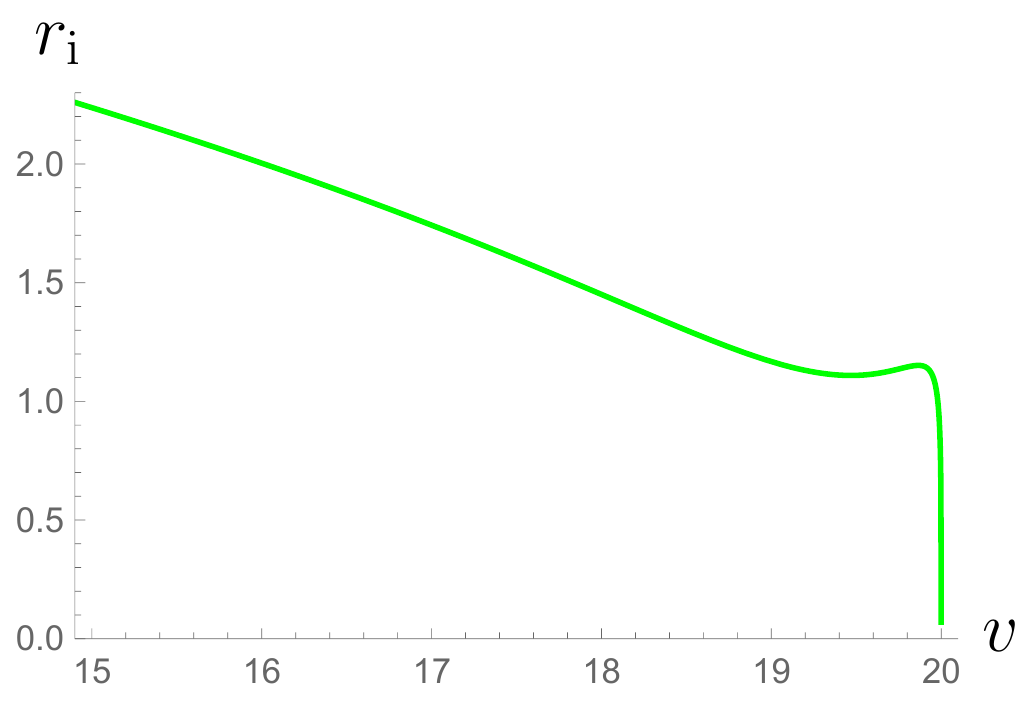}
	\includegraphics[scale=.5]{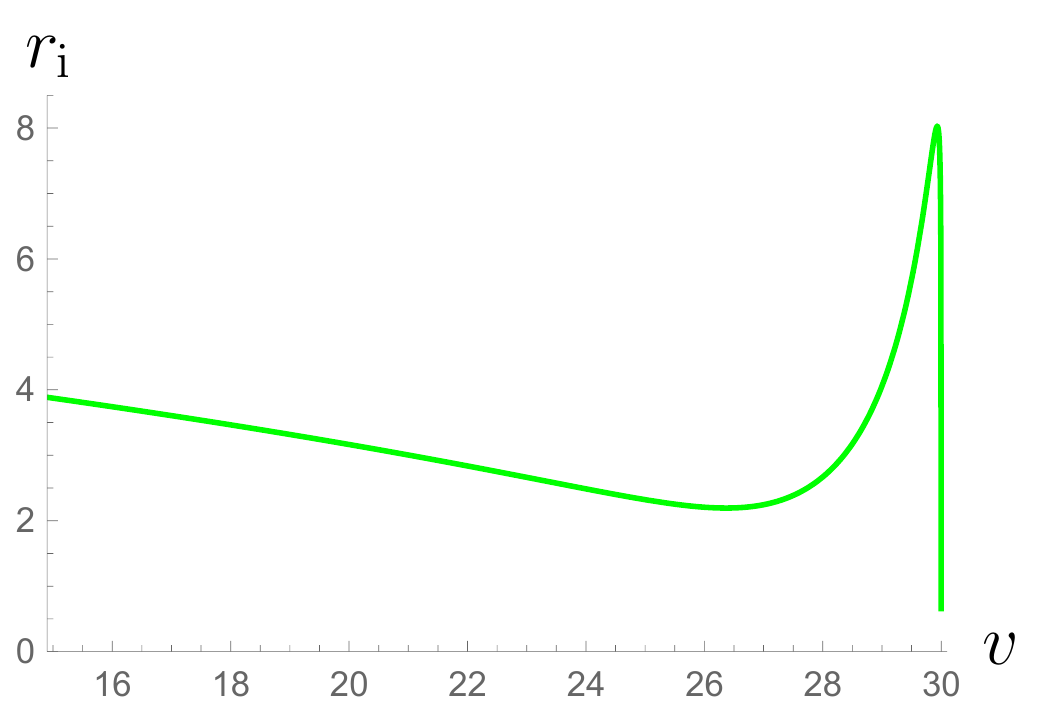}
	\caption{Plots of $r_{\rm i}(v)$ for a background given by \eqref{a1div} with $\lambda_0=1$ and $n=1/2$, with $v_{\rm s}=20$ on the left and with $v_{\rm s}=30$ on the right (starting from zero perturbation at $v=0$). Same units as above.}
	\label{bounce12}
\end{figure}

Indeed, if we integrate \eqref{sc}, which contains these initial tendencies along with their extrapolation to the regime in which semiclassical effects dominate, we get two different types of solutions:
\begin{enumerate}[label=(\roman*)]
	\item For $n\ge1$ the semiclassical backreaction always ends up overcoming the contribution of the classical background, resulting in a bounce in the position of the inner horizon, as shown in fig. \ref{bounce2}. We note that the final stages of the Oppenheimer-Snyder collapse correspond to a value $n=1$, as shown in \eqref{OSk}.
	\item For $n<1$, semiclassical backreaction can accumulate and lead to an initial bounce for large enough values of $v_{\rm s}$, as can be seen in fig. \ref{bounce12}. Such a bounce indicates that the collapsing behaviour of the classical matter has been temporarily counteracted, and may subsequently be inverted, making the trapped region disappear completely. However, in our extrapolation, due to the fact that the trapped region is not bounded from above, semiclassical effects eventually lose out and a singularity forms.
\end{enumerate}

We note that the time at which the bounce occurs depends on when this exponential overcomes the Planck length suppression, which generally occurs well before either the surface gravity or the radial position of the inner horizon get close to the Planck scale, as can be checked by inspecting figures \ref{bounce2} and \ref{bounce12}. On the one hand, this allows us to see that we have not obtained an unnaturally large result due to using the Polyakov approximation for the RSET [this already being obvious from the origin of the exponential in e.g. \eqref{alphastat}]. On the other hand, it is an indication that such dynamics can be treated semiclassically.\par
In summary, horizon-related semiclassical effects during gravitational collapse can only be avoided in this model if the classical trajectory of the inner horizon goes to zero quickly (small $v_{\rm s}$, not giving the exponential time to grow) and with a sharp peak at the end in $(v,r)$ coordinates ($n<1$, making the integral of $\lambda$ convergent). Otherwise, in a regime where semiclassical effects are dominated by a term like \eqref{Tvvin}, the collapse will tend to a halt, followed by a quick evaporation of the trapped region (or ``explosion") from the inside.\par
However, we remind the reader that  the accuracy of our approximation for the RSET \eqref{Tvvin} can break down when $\alpha$ becomes comparable to $\lambda$ at the scale of $r_{\rm i}$. Furthermore, in these dynamical scenarios it may become inaccurate even sooner if the inner horizon reaches a region sufficiently close to the origin to cross paths with light rays which have explored the core of the forming black hole, i.e. when it steps out of the left shaded region in fig. \ref{f2}. Then the precise structure of this core must be specified in order to calculate the RSET. Therefore, although this behaviour is the natural extension of our approximation, we cannot claim with certainty that it represents the complete semiclassical dynamics of gravitational collapse. However, it is a very suggestive possibility.\par

\section{Conclusions}

In this work we produce a bare-bones picture of semiclassical backreaction on black-hole spacetimes which have an inner horizon in addition to an outer one. We construct a simple toy model of a spherically-symmetric geometry in which a regular black hole forms, and look at the perturbations caused by the RSET around both the inner and outer apparent horizons. We find that treating these perturbations locally yields analytical results, and we obtain a clear picture of the initial tendencies of this double-horizon structure to evaporate. For the RSET we use a massless scalar field and apply the Polyakov approximation, which consists of dimensionally reducing the angular degrees of freedom, calculating the RSET in 1+1 dimensions and then returning to 3+1 dimensions with an approximation \cite{Fabbri2005}.\par
At the external horizon, the RSET provides an ingoing flux of negative energy (in accordance with the results of \cite{DFU}). Backreaction from this flux generates a small perturbation which tends toward evaporating the horizon. For a Schwarzschild geometry this perturbation initially grows proportionally to the advanced Eddington-Finkelstein time coordinate $v$, with a proportionality constant which includes the Planck length squared, i.e. a slow evaporation. For a background which is not a vacuum spacetime (e.g. Schwarzschild-dS, Schwarzschild-AdS, regular black holes), the modified relation between the radial position of the horizon and its surface gravity results in a somewhat different behaviour for the perturbation: if the surface gravity is larger than it would be for a Schwarzschild black hole of the same size, the evaporation is initially quicker than in Schwarzschild, but then has a tendency to slow down, and vice versa if the surface gravity is smaller.\par
At the inner horizon, the RSET again gives us a negative ingoing flux. The backreaction in this case again results in a reduction of the size of the trapped region, i.e. the inner horizon moves outward. Most importantly, this movement has an overall initial tendency to be much quicker that the evaporation of the outer horizon. This calculation of first order perturbations, if taken as indicative of the qualitative nature of the long-term evolution, strongly suggests a revised picture for evaporation: instead of the outer horizon slowly moving in and eventually revealing the core of the black hole, if an inner horizon is present, the trapped region may evaporate more quickly from the inside out. For regular black holes, this coincides with the picture described in \cite{Carballo-Rubio2019,Carballo-Rubio2019b}, which was motivated heuristically by the existence of mass inflation due to classical perturbations \cite{Carballo-Rubio2018}, although without an explicit discussion of the associated backreaction. Our results here show that the backreaction from semiclassical effects contains the seeds that may lead to a realization of this kind of picture.\par
In light of these results we extend our background geometries to include dynamical horizons. On the one hand, we do so in order to obtain a better approximation to the complete self-consistent semiclassical solutions which start from a static background. On the other, we are also interested in the backreaction around the dynamical inner horizon in models of black hole formation (e.g. Oppenheimer-Snyder collapse \cite{Oppenheimer1939}), where the trapped region first appears close to the eventual outer horizon, and its inner bound quickly moves inward, either reaching the origin (and forming a singularity) or tending to a halt before it (and leaving either a regular centre or one with a timelike singularity).\par
Through analysing these additional geometries we indeed obtain approximations for the self-consistent solutions in both static and dynamical backgrounds. Though the range of validity of these approximations is limited, they at least show us the initial tendency of the evolution quite clearly. We find that the semiclassical tendency to evaporate the trapped region remains even in dynamical backgrounds, though whether this can significantly affect the evolution of the geometry varies on a case-by-case basis. Most notably, we find that in many cases in which the background dynamics would make the inner horizon reach the origin (Oppenheimer-Snyder-type collapse), there is a tendency for semiclassical effects to become dominant before this occurs and bounce the horizon back outward. Although the bounce itself occurs in most cases outside the range of validity of the approximation we use for calculating the RSET, the way this result depends on the divergent tendency of the surface gravity is very suggestive of it being a generic property of geometries of this type.\par
Given that all astrophysical black holes are expected to have an inner horizon, be it a dynamical one during their formation or perturbation, or a stationary one left behind due to angular momentum (and/or electric charge), our results at the very least indicate that horizon-related semiclassical effects should never be overlooked when analysing their formation and evolution.

\section*{Acknowledgements}
Financial support was provided by the Spanish Government through the projects FIS2017-86497-C2-1-P, FIS2017-86497-C2-2-P (with FEDER contribution), FIS2016-78859-P (AEI/ FEDER,UE), and by the Junta de Andalucía through the project FQM219. VB is funded by the Spanish Government fellowship FPU17/04471. CB acknowledges financial support from the State Agency for Research of the Spanish MCIU through the ``Center of Excellence Severo Ochoa'' award to the Instituto de Astrof\'{\i}sica de Andaluc\'{\i}a (SEV-2017-0709).

\section*{References}

\nocite{*}
\bibliography{Bibliografia}
\bibliographystyle{ieeetr}

\end{document}